\documentclass[%
preprint
 amsmath,amssymb,
 aps,
 twocolumn,
 prb,
floatfix,
]{revtex4-2}

\usepackage{graphicx}% Include figure files
\usepackage{dcolumn}% Align table columns on decimal point
\usepackage{bm}% bold math
\usepackage{array} % for advanced table features
\usepackage{chemformula} % Formula subscripts using \ch{}
\usepackage[T1]{fontenc} % Use modern font encodings
\usepackage[version=3]{mhchem}
\usepackage{stfloats} 
\usepackage{graphicx} 
\usepackage[most]{tcolorbox}
\usepackage{placeins}
\usepackage{tcolorbox}
\usepackage{lastpage}
\usepackage[format=plain,justification=justified,singlelinecheck=false,font={stretch=1.125,small,sf},labelfont=bf,labelsep=space]{caption}
\usepackage{float}
\usepackage{fancyhdr}
\usepackage{tabularx} 
\usepackage{threeparttable}
\usepackage{fnpos}
\usepackage[english]{babel}
\usepackage{array}
\usepackage{droidsans}
\usepackage[title]{appendix}
\usepackage{charter}
\usepackage[T1]{fontenc}
\usepackage[usenames,dvipsnames]{xcolor}
\usepackage{setspace}
\usepackage[compact]{titlesec}
\usepackage{subcaption}
\usepackage{booktabs}
\definecolor{cream}{RGB}{222,217,201}
\usepackage{xcolor}
\usepackage{braket}
\begin{document}

\title{Computing Reaction and Activation Energies of Pericyclic Reactions using a Symmetry-Adapted VQE Algorithm}

% \preprint{Qauntum simulation of organic reactions on quantum computer}% Force line breaks with \\

\author{Maitreyee Sarkar$^{1}$}
% \altaffiliation[Also at ]{Physics Department, XYZ University.}
\email{sarkar.5@iitj.ac.in}
\author{Manikandan Paranjothy$^{2}$}%
\email{pmanikandan@iitj.ac.in}
\author{Atul Kumar$^{1}$}%
\email{atulk@iitj.ac.in}
\affiliation{%
$^{1}$ Quantum Information and Computation Lab, Department of Chemistry, Indian Institute of Technology Jodhpur, Rajasthan, India, 342030 %\textbackslash\textbackslash
}%
\affiliation{%
$^{2}$ Chemical Dynamics Research Group, Department of Chemistry, Indian Institute of Technology Jodhpur, Rajasthan, India, 342030 %\textbackslash\textbackslash
}%

\begin{abstract}
Pericyclic reactions provide stringent tests for quantum simulations because their mechanisms are governed by orbital symmetry and involve correlated transition states. In this work, we employ the variational quantum eigensolver (VQE) combined with a previously established symmetry-guided active-space selection protocol based on symmetry-matched fractions (SMF-VQE) to simulate Diels–Alder and Alder–ene reactions in complex systems involving extended $\pi$-conjugation and multiple bonding. Although absolute electronic energies obtained from the current protocol exhibit significant deviations from the values computed using CCSD method, the symmetry-guided active spaces yield substantial cancellation of deviations in the energy differences. As a result, reaction energies are predicted with error (relative to CCSD) less than 1 kcal/mol, while activation energies are reproduced within about 5–6 kcal/mol. The symmetry-guided protocol also reduces the large combinatorial space of active-space choices to a single symmetry-consistent selection for each reaction.
\end{abstract}

%Use showkeys class option if keyword
\maketitle

%\tableofcontents

\section{\label{sec:level1} Introduction}

Quantum computing has generated considerable interest as a framework to benchmark, complement, and potentially extend conventional electronic structure calculations for chemically relevant problems, particularly where electron correlation and Hilbert-space complexity becomes challenging \cite{cao2019quantum, mcardle2020quantum, lanyon2010towards, claudino2022basics, kais2014introduction, fano2019quantum, kassal2011simulating, dirac1929quantum}. Within the Noisy Intermediate-Scale Quantum (NISQ) era \cite{preskill2018quantum}, hybrid quantum-classical approaches provide opportunities to investigate how quantum algorithms may facilitate such calculations, where feasible within available quantum resources, while also serving as platforms to assess emerging methodologies for molecular simulations. Among these approaches, the variational quantum eigensolver (VQE) has attracted significant attention as a variational framework for estimating molecular ground-state energies using parametrized quantum circuits coupled with classical optimization \cite{armaos2020computational, feynman2018simulating, aspuru2005simulated, peruzzo2014variational}. Although present NISQ implementations are limited by finite resources and noise, they provide a useful setting for exploring active-space strategies, electron correlation, and symmetry considerations relevant to quantum chemical calculations. \par
Pericyclic reactions provide particularly suitable benchmark systems for such investigations because their mechanisms are governed by orbital symmetry and concerted electronic rearrangements. Their reactivity is classically understood through orbital symmetry conservation principles embodied in the Woodward-Hoffmann framework, while their transition states often involve subtle correlation effects associated with simultaneous bond formation and bond breaking. These characteristics make them chemically meaningful test cases for assessing quantum algorithms beyond single-molecule ground-state calculations. These reactions also provide a more demanding test than simpler molecular benchmarks because they involve larger active spaces, extended $\pi$ interactions, and in some cases multiple bonding. Such features make them appropriate systems for assessing symmetry-guided quantum simulations in chemically more complex molecular settings. In this context, symmetry can play a dual role: it is central to the chemistry of pericyclic reactions and can also guide reduced yet chemically consistent active spaces in VQE calculations. The present work examines Diels-Alder and Alder-Ene reactions, including their transition states, as benchmark cases for assessing symmetry-guided quantum simulations in chemically more complex systems involving larger active spaces and multiple bonding. To this end, we employ a previously reported \cite{Sarkar2026} symmetry-matched fractions (SMFs) as the active-space selection criterion used in the present calculations and assess whether such symmetry-consistent selection enables reliable evaluation of reaction and activation energetics despite deviations in absolute VQE energies. \par
A recent quantum simulation of Diels-Alder reactivity by Liepuoniute \textit{et al.} employed entanglement forging, quantum subspace expansion, and perturbative corrections to study activation barriers using chemically selected active spaces \cite{D4CP01314J}. The present work differs in both focus and methodology. Rather than emphasizing algorithmic implementation on quantum hardware, we investigate Diels-Alder and Alder-Ene reactions as benchmark systems for assessing symmetry-guided quantum simulations in larger and chemically more demanding systems, including transition states. Unlike active spaces chosen primarily from chemical intuition, the present work employs a symmetry-guided criterion applied consistently across all species along the reaction coordinate and analyzes how this reduces the combinatorial choices of active spaces while retaining accurate reaction energetics. \par
Using CCSD as the classical reference and VQE calculations within this symmetry-guided framework, we examine whether systematic cancellation of errors in energy differences enables reliable estimation of reaction and activation energetics, even when absolute VQE energies exhibit significant deviations with respect to CCSD predictions. The results suggest that chemically complex pericyclic reactions provide valuable benchmark systems for assessing symmetry-guided quantum simulations within present resource constraints.

\section{Theoretical Overview}
\subsection{Variational Quantum Eigensolver Algorithm}
%new fig
% \begin{figure}[h!]
% \centering
% \includegraphics[width=0.5\textwidth]{vqe.png}
%  \caption{VQE}
% \end{figure}
The variational quantum eigensolver (VQE) is a hybrid quantum-classical algorithm used to estimate molecular ground-state energies \cite{alteg2022study} based on the variational principle \cite{mcquarrie1997physical}, where a parametrized trial wavefunction is represented as a quantum circuit. The molecular Hamiltonian is transformed into a qubit Hamiltonian using simultaneous second quantization and fermion-to-qubit mapping \cite{alteg2022studyadaptativederivativeassemblepseudotrotter}. In this work, the trial wavefunction is constructed using the unitary coupled cluster singles and doubles (UCCSD) ansatz, and variational parameters are optimized classically \cite{singh2023benchmarking}. UCCSD is particularly suitable in the present context because its excitation operators provide a natural connection between the variational ansatz and the symmetry-guided excitation analysis used in the present work.

\begin{figure}[h]
\centering
\begin{subfigure}{0.45\textwidth}
\centering
\includegraphics[width=1.1\textwidth]{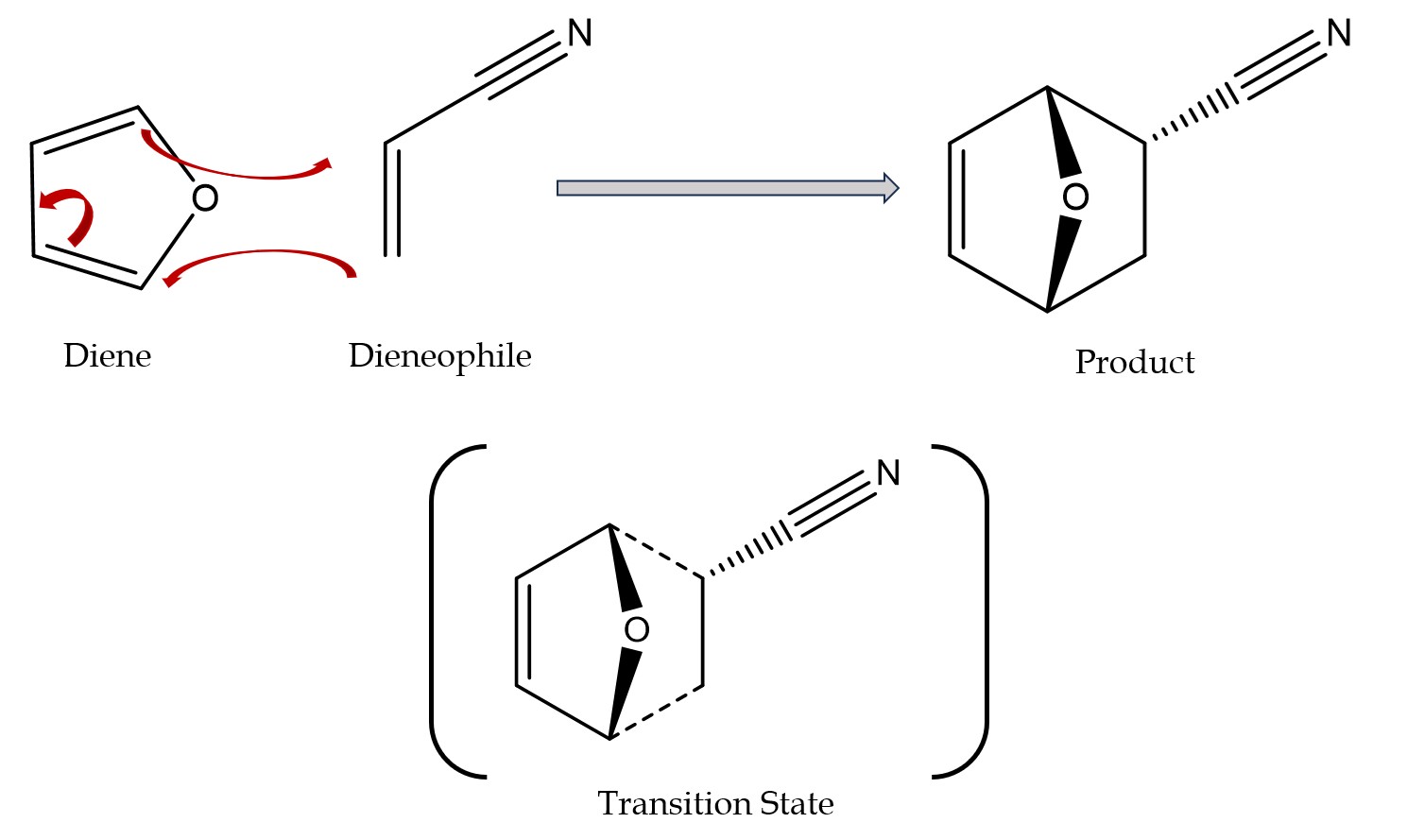}
\caption{Diels-Alder Reaction}
\label{da}
\end{subfigure}
\hfill
\begin{subfigure}{0.45\textwidth}
\centering
\includegraphics[width=1.1\textwidth]{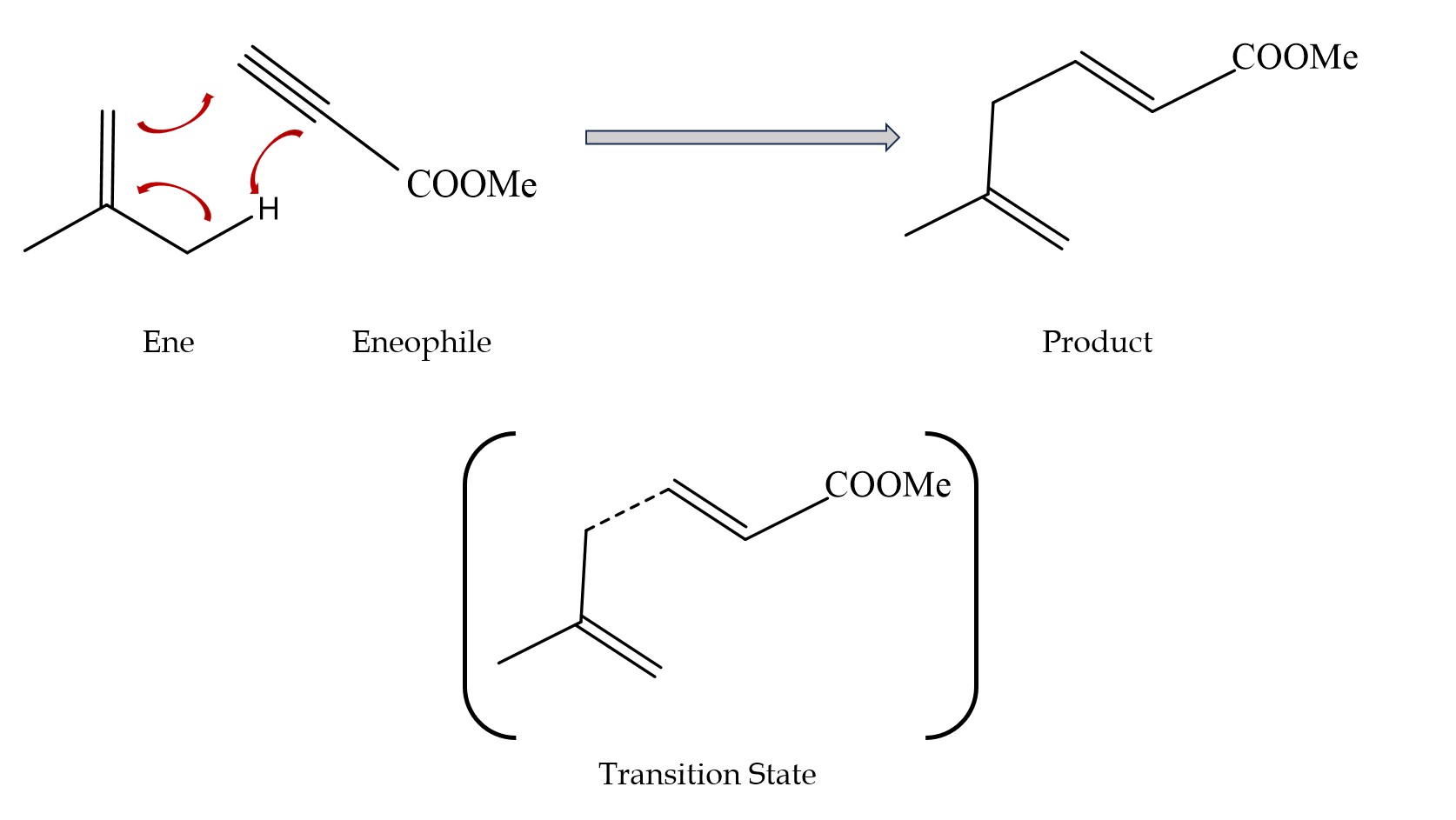}
\caption{Alder-Ene Reaction}
\label{ae}
\end{subfigure}
\caption{Diels-Alder and Alder-Ene reactions used in this study}
\label{da-ae}
\end{figure}

% %new fig box
% \begin{figure}[h!]
% \centering
% \begin{tcolorbox}[
%   enhanced,
%   width=0.45\textwidth,
%   colback=white!5!white,
%   colframe=violet!80!black,
%   boxrule=0.4pt,
%   title=Diels-Alder Reaction
% ]
% \includegraphics[width=\textwidth]{da.jpg}
% \label{da}
% \end{tcolorbox}
% \end{figure}

% % %new fig box
% \begin{figure}[h!]
% \centering
% \begin{tcolorbox}[
%   enhanced,
%   width=0.45\textwidth,
%   colback=white!5!white,
%   colframe=violet!80!black,
%   boxrule=0.4pt,
%   title=Alder-ene Reaction
% ]
% \includegraphics[width=\textwidth]{ae.jpg}
% \label{ae}
% \end{tcolorbox}
% \end{figure}

 % in preamble

% \begin{figure}[h!]
% \centering
% \begin{subfigure}{0.45\textwidth}
% \centering
% \begin{tcolorbox}[
%   enhanced,
%   width=\textwidth,
%   colback=white!5!white,
%   colframe=violet!80!black,
%   boxrule=0.4pt,
%   title=Diels-Alder Reaction
% ]
% \includegraphics[width=\textwidth]{da.jpg}
% \end{tcolorbox}
% \caption{}
% \label{da}
% \end{subfigure}
% \hfill
% \begin{subfigure}{0.45\textwidth}
% \centering
% \begin{tcolorbox}[
%   enhanced,
%   width=\textwidth,
%   colback=white!5!white,
%   colframe=violet!80!black,
%   boxrule=0.4pt,
%   title=Alder-ene Reaction
% ]
% \includegraphics[width=\textwidth]{ae.jpg}
% \end{tcolorbox}
% \caption{}
% \label{ae}
% \end{subfigure}
% \caption{Comparison of Diels–Alder and Alder–ene reactions}
% \end{figure}

\subsection{Pericyclic Reactions and Symmetry Considerations}
Pericyclic reactions are concerted transformations governed by orbital symmetry and cyclic electron rearrangements \cite{fleming2015pericyclic,epiotis1974theory}. Because bond formation and bond breaking occur simultaneously through correlated transition states, these reactions provide useful test cases for examining whether symmetry-guided quantum simulations can describe chemically relevant reaction pathways. The present reactions provide a more demanding test than simpler molecular benchmarks because they involve larger active spaces, extended $\pi$ interactions, and in some cases multiple bonding. These features make them particularly suitable for examining symmetry-guided active-space selection in quantum chemical calculations.
Among such reactions, Diels-Alder and Alder-Ene reactions serve as prototypical examples with well-defined orbital symmetry constraints and constitute the systems considered in this study.

\textit{Diels-Alder Reaction:} The Diels-Alder reaction considered here is a [4+2] cycloaddition involving concerted interaction between a conjugated diene and a dienophile through a six-membered cyclic transition state.

\textit{Alder-Ene Reaction:} The Alder-Ene reaction considered here proceeds through concerted allylic hydrogen transfer accompanied by new $\sigma$-bond formation and migration of a $\pi$ bond through a cyclic transition state.

\section{Methods}

\subsection{Classical Computation}
Classical electronic structure calculations for all reactants, products, and transition states were performed using the NWChem package \cite{valiev2010nwchem}. Geometry optimizations were performed at the density functional theory (DFT) M05-2X method using the cc-pVDZ basis set, and ground-state single-point energy calculations were carried out at the CCSD/STO-3G level utilizing the DFT optimized geometries. Reaction energies and activation energies were subsequently evaluated for both pericyclic reactions. For the symmetry-guided analysis, irreducible representations of molecular orbitals were determined by applying the highest available Abelian point-group symmetry for each species.

\subsection{Quantum Computation}
VQE calculations were performed using Qiskit 1.1.1 \cite{qiskit2024} and Qiskit Nature 0.7.2 \cite{qiskit_nature_2023} with the STO-3G basis and a UCCSD ansatz. Electronic Hamiltonians were mapped to qubit operators using parity mapping with tapering to reduce the number of qubits. \par
The present systems involve extended $\pi$ interactions, multiple bonding, and transition states, which substantially increase the size of the active spaces, the number of excitation operators, and the associated quantum resources required within the UCCSD framework. In order to examine symmetry-guided active-space selection for chemically more complex pericyclic systems within present computational and quantum resource constraints, all calculations were performed using the STO-3G basis set. Geometry optimizations were carried out at the M05-2X level, while CCSD single point energies (computed using DFT geometries) were used as classical reference energies. All calculations employed the Aer noiseless simulator in order to isolate the role of symmetry-guided active-space selection and correlated excitation analysis from hardware-induced noise effects. Variational optimization was carried out using the sequential least squares programming (SLSQP) algorithm.

\subsection{Symmetry-Guided Active-Space Selection}
The choice of active space strongly influences the quality and consistency of VQE energetics, particularly when comparing reactants, products, and transition states within a common reaction pathway. In the present work, active spaces were selected using a symmetry-guided criterion based on symmetry-matched fractions (SMFs), following the group-theoretic framework employed in our earlier studies on reaction and ring-strain energetics \cite{cotton1991chemical,mirman1995group,butler2012point, Sarkar2026,roy2026reaction}. The objective here is not only to reduce the number of candidate active spaces, but also to maintain a symmetry-consistent treatment of electron correlation across all species involved in a reaction. \par

For each molecule, multiple candidate active spaces ranging from $(2e,3o)$ to $(8e,8o)$ were examined. Within a given active space, all possible excitation operators generated through the UCCSD framework were classified according to their irreducible representations relative to the Hartree-Fock reference state. Excitations transforming according to the same irreducible representation as the reference state are expected to contribute more directly to the correlated ground-state description. Based on this consideration, the symmetry-matched fraction (SMF) was defined as
\begin{equation}
\mathrm{SMF} =
\frac{N_{\text{same-irrep}}}
     {N_{\text{total-excitations}}}
\times 100
\label{eq:smf}
\end{equation}
where $N_{\mathrm{same-irrep}}$ denotes the number of excitation operators having the same irreducible representation as the reference state, and $N_{\mathrm{total-excitations}}$ represents the total number of excitations generated within the chosen active space \cite{Sarkar2026}. The numerator in the above expression will be referred to as the symmetry-matched value (SMV). For each reaction, the active spaces of reactants, products, and transition states were compared on the basis of their SMF values. Active spaces with the largest SMF values were selected to maintain symmetry consistency across the reaction coordinate while retaining a manageable variational space for the VQE calculations. \par
\begin{table}[]
\caption{\label{variable-da} Variables for Diels-Alder Reaction}
\begin{ruledtabular}
\begin{tabular}{lccr}
\textrm{Molecule}&
\textrm{Symmetry}&
\multicolumn{1}{c}{\textrm{Pool}}&
\textrm{min\_orb, max\_orb}\\
\midrule
Diene & $C_2v$ & 4o, 4v & 3,6\\
dienophile  & $C_s$ & 4o, 4v & 3,6\\
Product & $C_1$ & 4o, 2v & 3,6\\
TS & $C_1$ & 4o, 5v & 3,8\\
\end{tabular}
\end{ruledtabular}
\end{table}

\begin{table}[]
\caption{\label{variable-ae} Variables for Alder-ene Reaction}
\begin{ruledtabular}
\begin{tabular}{lccr}
\textrm{Molecule}&
\textrm{Symmetry}&
\multicolumn{1}{c}{\textrm{Pool}}&
\textrm{min\_orb, max\_orb}\\
\colrule
Ene & $C_1$ & 4o, 2v & 3,6\\
enophile & $C_s$ & 4o, 4v & 3,6\\
Product & $C_1$ & 4o, 2v & 3,6\\
TS & $C_1$ & 4o, 5v & 3,8\\
\end{tabular}
\end{ruledtabular}
\end{table}
As organic molecules generally possess lower symmetry, the systems considered in this work belong primarily to the $C_{2v}$, $C_s$, and $C_1$ point groups. For species with $C_{2v}$ and $C_s$ symmetry, an eight-orbital pool consisting of four occupied and four virtual orbitals was employed. For $C_1$ species appearing among reactants and products, a reduced six-orbital pool consisting of four occupied and two virtual orbitals was used. In $C_1$ symmetry, all excitations belong to the same irreducible representation, resulting in identical SMF values for all active spaces. In such cases, the SMV was used to guide active-space selection instead of the SMF. \par

Transition states were treated separately from the remaining $C_1$ species because they exhibit near-degeneracy and partial bond-breaking characteristics. To account for these effects, larger orbital pools consisting of four occupied and five virtual orbitals were considered for the transition states of both reactions. To maintain consistency across all calculations, common orbital pools and active-space ranges were employed for each reaction. The details of the orbital pools, point-group symmetries, and active-space ranges used in the present work are summarized in Tables \ref{variable-da} and \ref{variable-ae}.

\section{Results and Discussion}
\begin{table}[h]
\caption{\label{comp-da} Reaction and Activation energies (in kcal/mol) for Diels-Alder Reaction and their differences}
\begin{ruledtabular}
\begin{tabular}{lccr}
\textrm{Energy}&
\textrm{CCSD }&
\textrm{SMF-VQE }&
\textrm{Diff.}\\
\colrule
Reaction & -34.23219 & -34.31457 & 0.08238\\
Activation & 24.34098 & 19.36314 & 4.97784\\
\end{tabular}
\end{ruledtabular}
\end{table}

%new fig
\begin{figure}[]
    \centering

    % ---------- Row 1 ----------
    \begin{subfigure}{0.4\textwidth}
        \centering
        \includegraphics[width=0.9\linewidth]{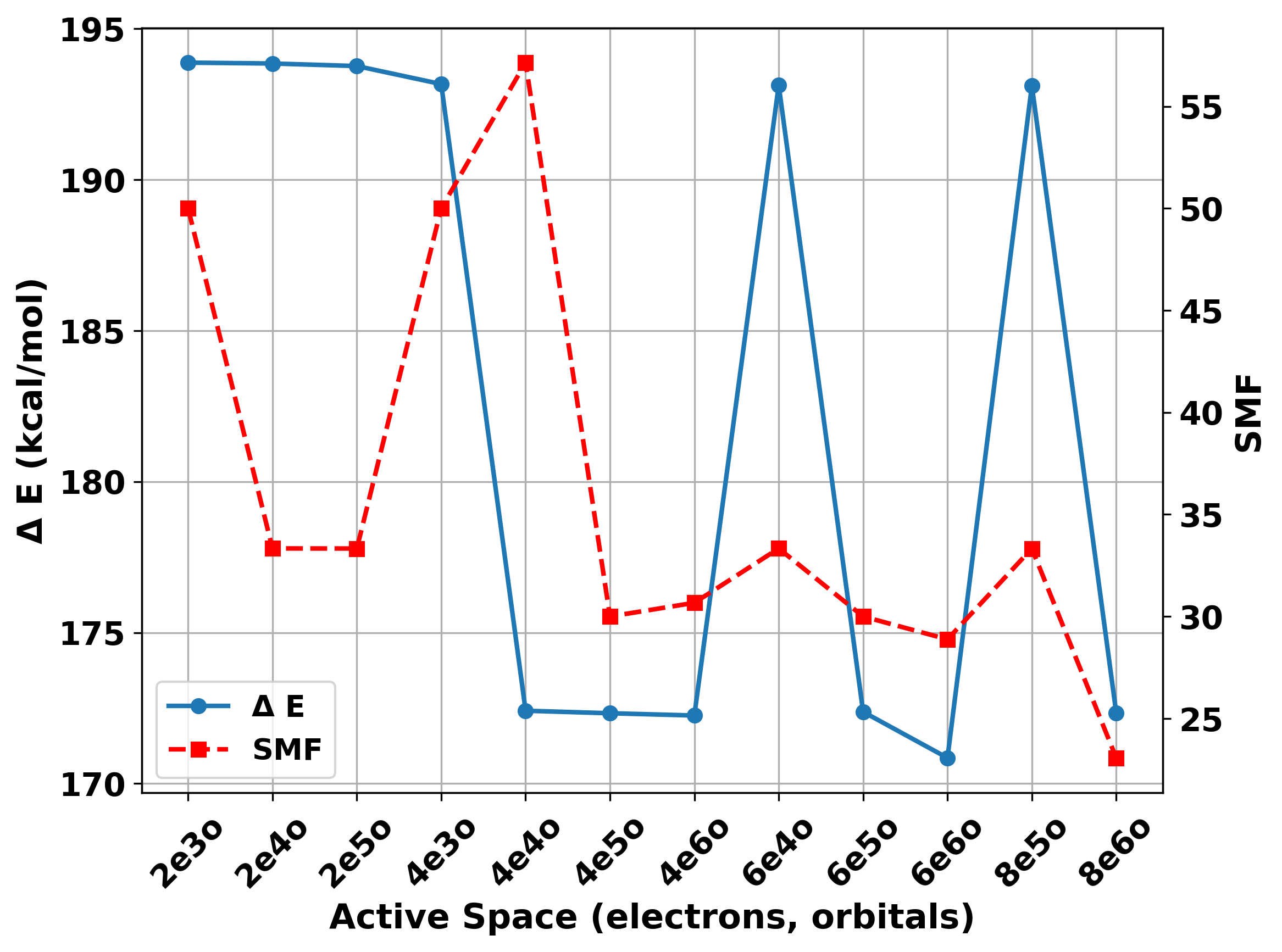}
        \caption{}
        \label{fig:hf_energy}
    \end{subfigure}
    % \hfill
    \begin{subfigure}{0.4\textwidth}
        \centering
        \includegraphics[width=0.9\linewidth]{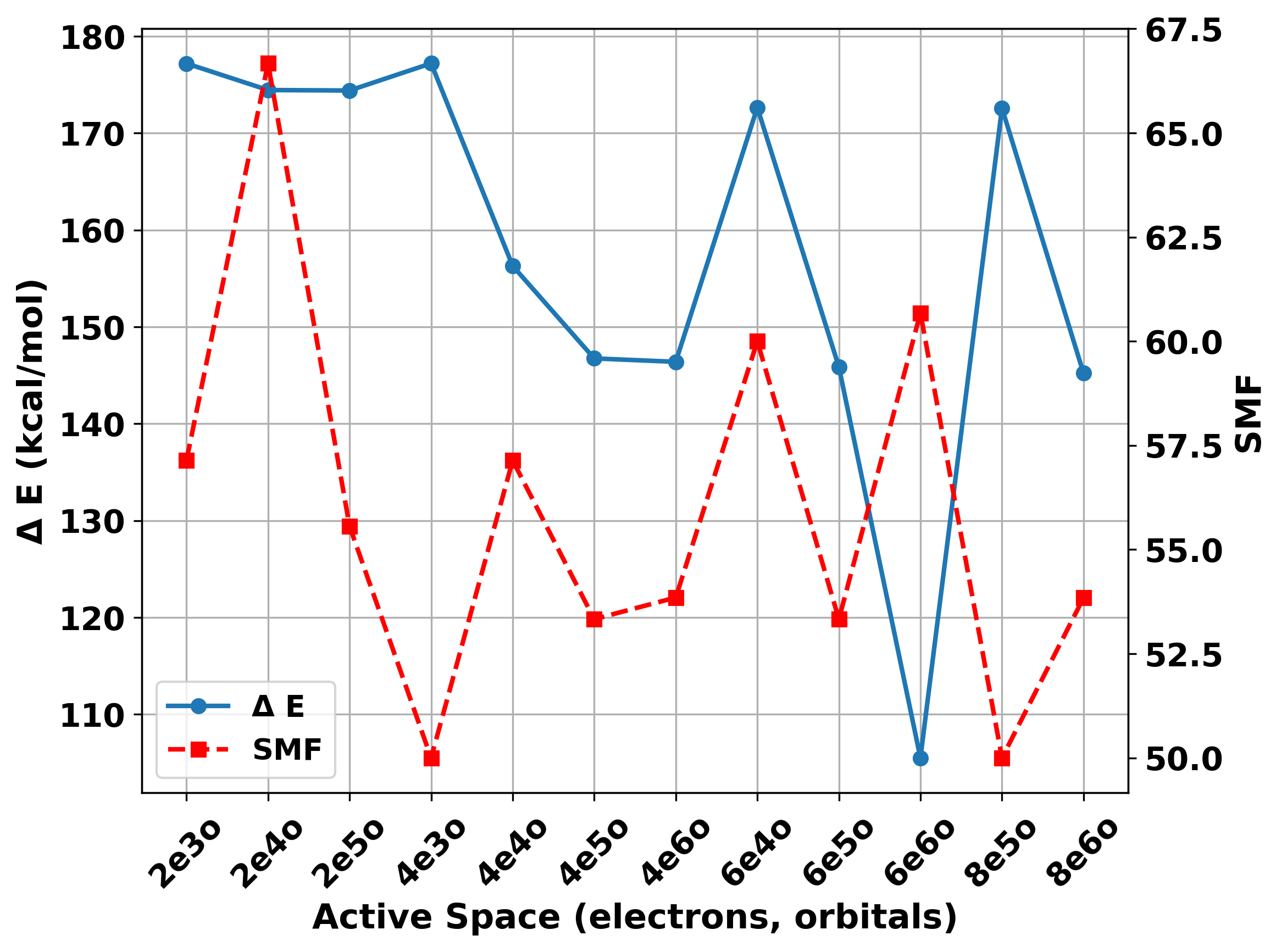}
        \caption{}
        \label{fig:hf_prob}
    \end{subfigure}

    % \vspace{0.3cm}

    % ---------- Row 2 ----------
    \begin{subfigure}{0.4\textwidth}
        \centering
        \includegraphics[width=0.9\linewidth]{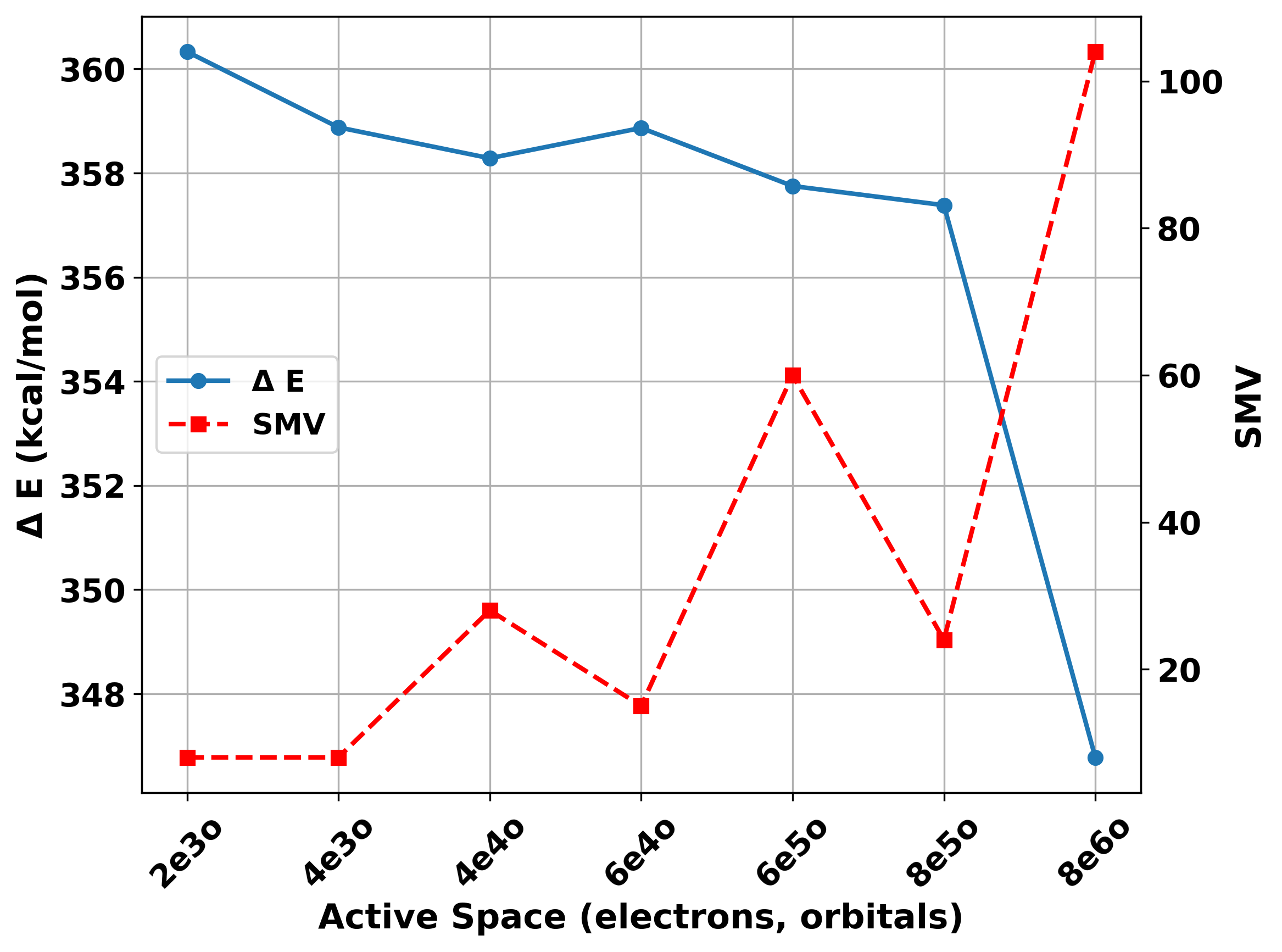}
        \caption{}
        \label{fig:h2_energy}
    \end{subfigure}
    % \hfill
    \begin{subfigure}{0.4\textwidth}
        \centering
        \includegraphics[width=0.9\linewidth]{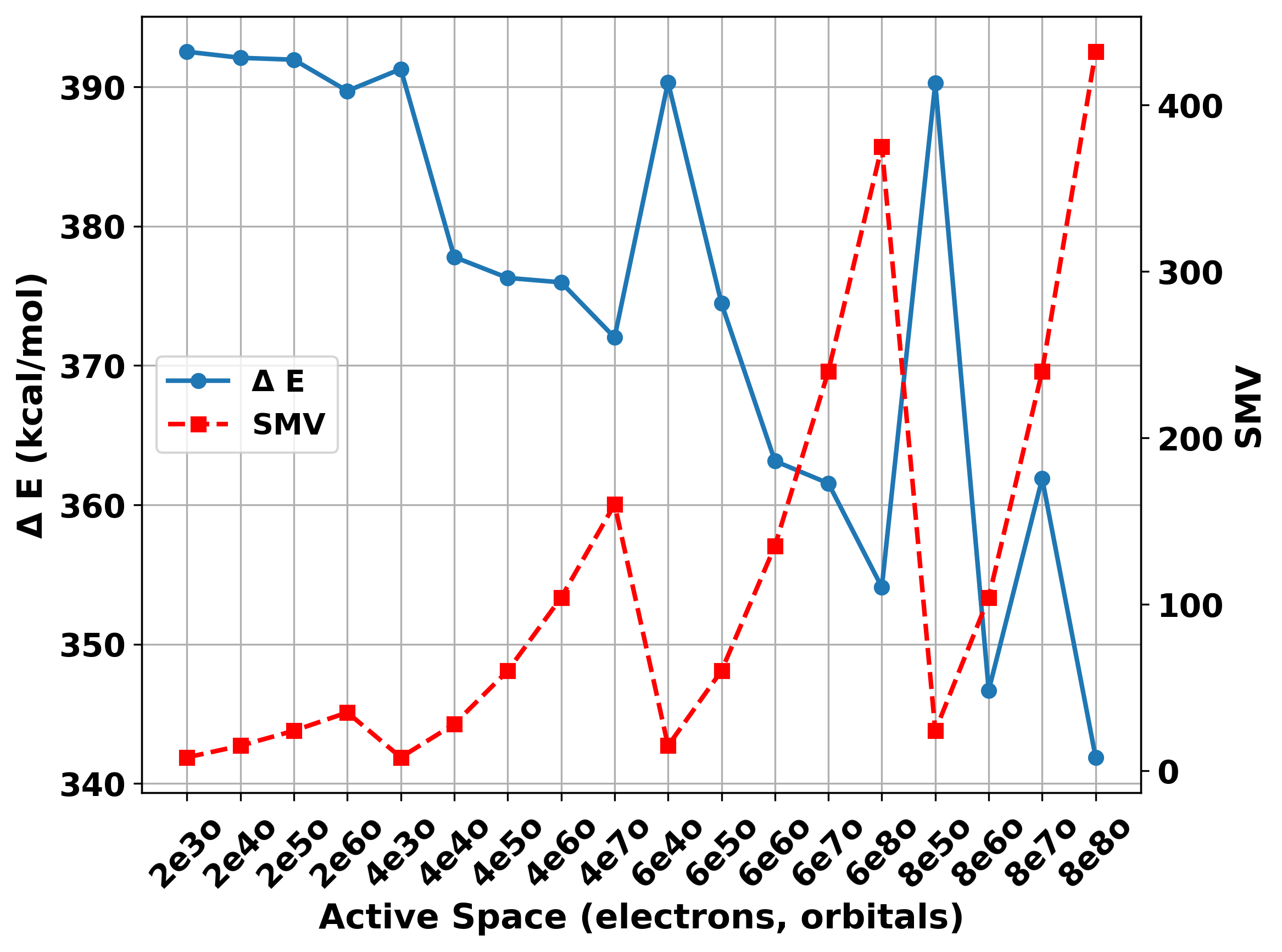}
        \caption{}
        \label{fig:h2_prob}
    \end{subfigure}

    \caption{Absolute Energy Error and SMF/SMV values for (a) diene, (b) dienophile, (c) product, and (d) transition state.}
    \label{main}
\end{figure}

\subsection{Diels-Alder Reaction}
We chose the Diels-Alder reaction in Fig.\ref{da-ae}(a) as it has furan as diene with oxygen (-O) as an electron donating group (EDG) because it's lone pair increases electron density of the conjugated double bonds and the dienophile contains cyano (-CN) group as an electron withdrawing group (EWG) which lowers the lowest unoccupied molecular orbital (LUMO) and makes the transition state and product more viable and stable. The diene possesses $C_{2v}$ symmetry and the dienophile possesses $C_s$ symmetry, as per table \ref{variable-da}, the active spaces are generated from a pool of four occupied and four vacant orbitals, and the active spaces generated will contain a maximum of six orbitals and a minimum 3 orbitals, resulting in 12 candidate active spaces for each species. The product state belongs to the $C_1$ point group and therefore, as per table \ref{variable-da}, the active spaces are generated from a pool of four occupied and two vacant orbitals, and the active spaces generated will contain a maximum of six orbitals and a minimum of three orbitals, resulting in 7 candidate active spaces. The transition state belongs to the $C_1$ point group and therefore, as per table \ref{variable-da}, the active spaces are generated from a pool of four occupied and five vacant orbitals, and the active spaces generated will contain a maximum of eight orbitals and a minimum of three orbitals, resulting in 18 candidate active spaces. The VQE calculations were performed for all candidate active spaces of the diene, dienophile, product, and transition state species involved in the Diels-Alder reaction. The SMF and SMV values were calculated as suggested in Eq. \ref{eq:smf}. Figure \ref{main} presents all of these active spaces and the absolute energy errors of all VQE calculations relative to CCSD, together with the corresponding SMF values for the diene and dienophile and SMV values for the product and transition state. Absolute energies are provided in the Supporting Information. It can be seen from  Figures \ref{main} (a)-(d) that when the number of electrons in the active spaces remains the same and the number of orbitals increases, a partial reduction in the absolute energy deviations happens for several active-space sequences with increasing orbital number, although the overall behavior remains non-monotonic. The electronic energies obtained from VQE exhibit substantial deviations from CCSD for all species, with average absolute errors of approximately 180, 140, 350, and 370 kcal/mol for the diene, dienophile, product, and transition state, respectively. However, the symmetry-guided active spaces selected through the SMF/SMV criterion yield substantially improved agreement in reaction and activation energetics due to cancellation of deviations in the energy differences. \par
The reaction energy is computed as
\begin{equation}
\Delta E =\sum_i E_i(\mathrm{products})-\sum_j E_j(\mathrm{reactants}) \label{eq:rxn}
\end{equation} 
And the energy values of the active spaces for reactants and products with maximum SMF/SMV are used. The activation energy is computed as
\begin{equation}
\Delta E =E(\mathrm{TS}) - \sum_i E_i(\mathrm{reactants}) \label{eq:actn}
\end{equation} 
 The active spaces AS(4e,4o), AS(2e,4o), AS(8e,6o), and AS(8e,8o) were found to possess the largest SMF or SMV values for the diene, dienophile, product, and transition state, respectively. These active spaces were therefore selected for evaluating reaction and activation energetics within the symmetry-guided framework.  Although 1008 possible combinations of reaction energies and 2592 possible combinations of activation energies can be generated from all active-space combinations, the symmetry-guided criterion reduces these possibilities to a single symmetry-consistent combination for each quantity. Figure \ref{comb} illustrates this reduction in the number of possible combinations. \par

\begin{figure}[h]
    \centering
    \includegraphics[width=0.9\linewidth]{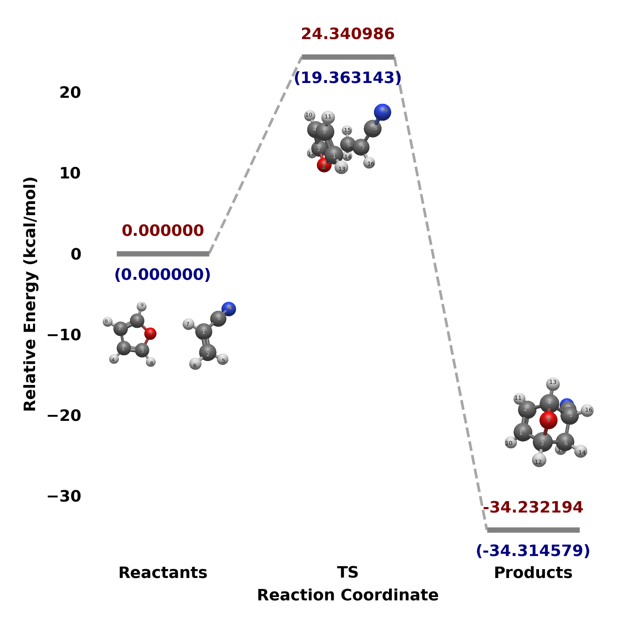}
    \caption{Comparison of PES for CCSD (red) and SMF-VQE (blue and inside bracket) for Diels-Alder Reaction}
    \label{pes1}
\end{figure}

The reaction and activation energies obtained from the selected SMF-guided active spaces are summarized in Table \ref{comp-da}. Despite the large deviations in absolute ground-state energies, substantial cancellation of errors occurs in the energy differences. The reaction energy differs from the CCSD value by only 0.08 kcal/mol, while the activation energy differs by approximately 5 kcal/mol, which can be seen in Figure \ref{pes1}. These results indicate that symmetry-consistent active-space selection can recover chemically meaningful reaction energetics even when the absolute VQE energies remain significantly shifted relative to CCSD.

% %new fig
\begin{figure}[]
    \centering

    % ---------- Row 1 ----------
    \begin{subfigure}{0.4\textwidth}
        \centering
        \includegraphics[width=0.9\linewidth]{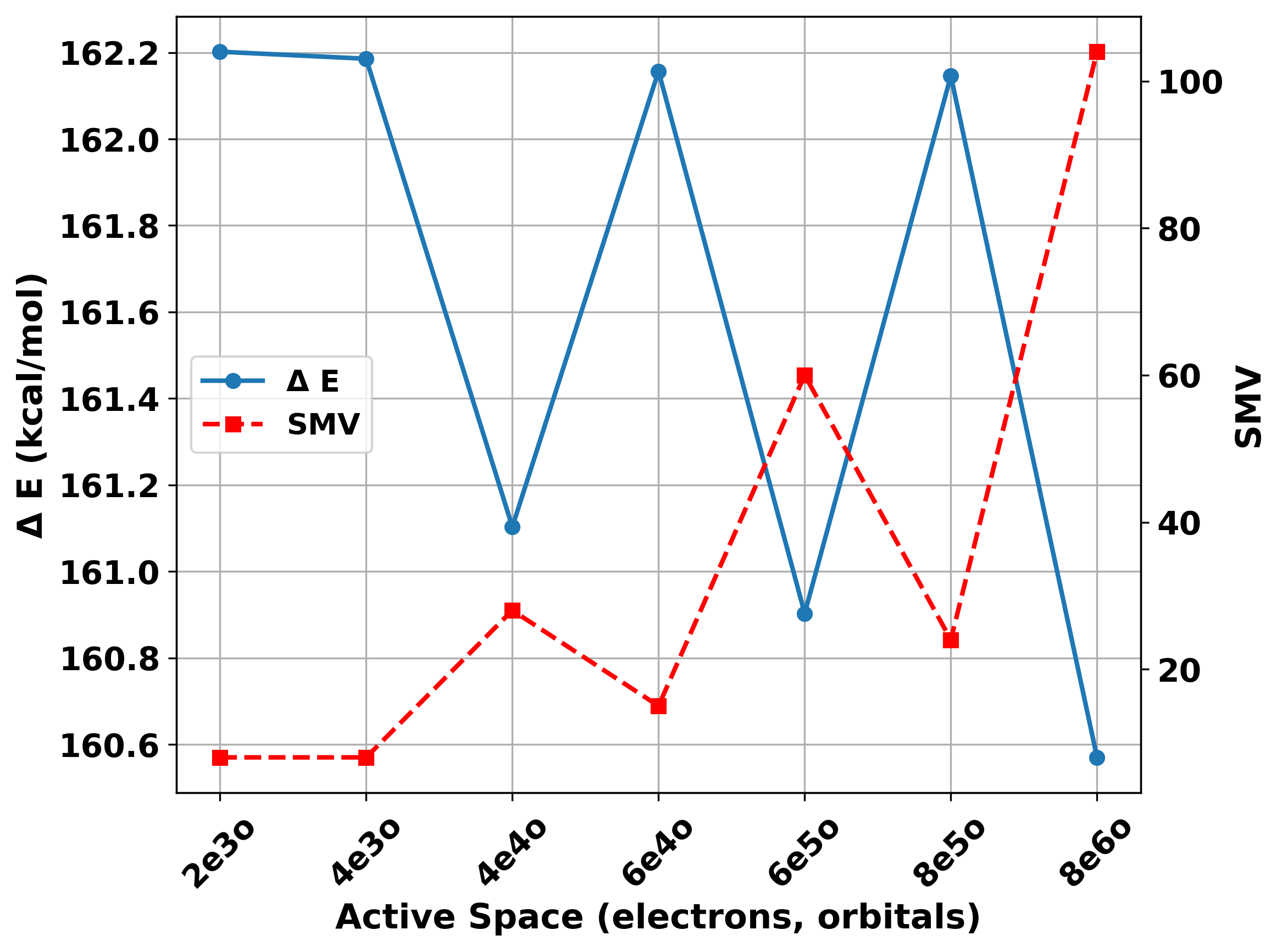}
        \caption{}
        \label{fig:hf_energy}
    \end{subfigure}
    % \hfill
    \begin{subfigure}{0.4\textwidth}
        \centering
        \includegraphics[width=0.9\linewidth]{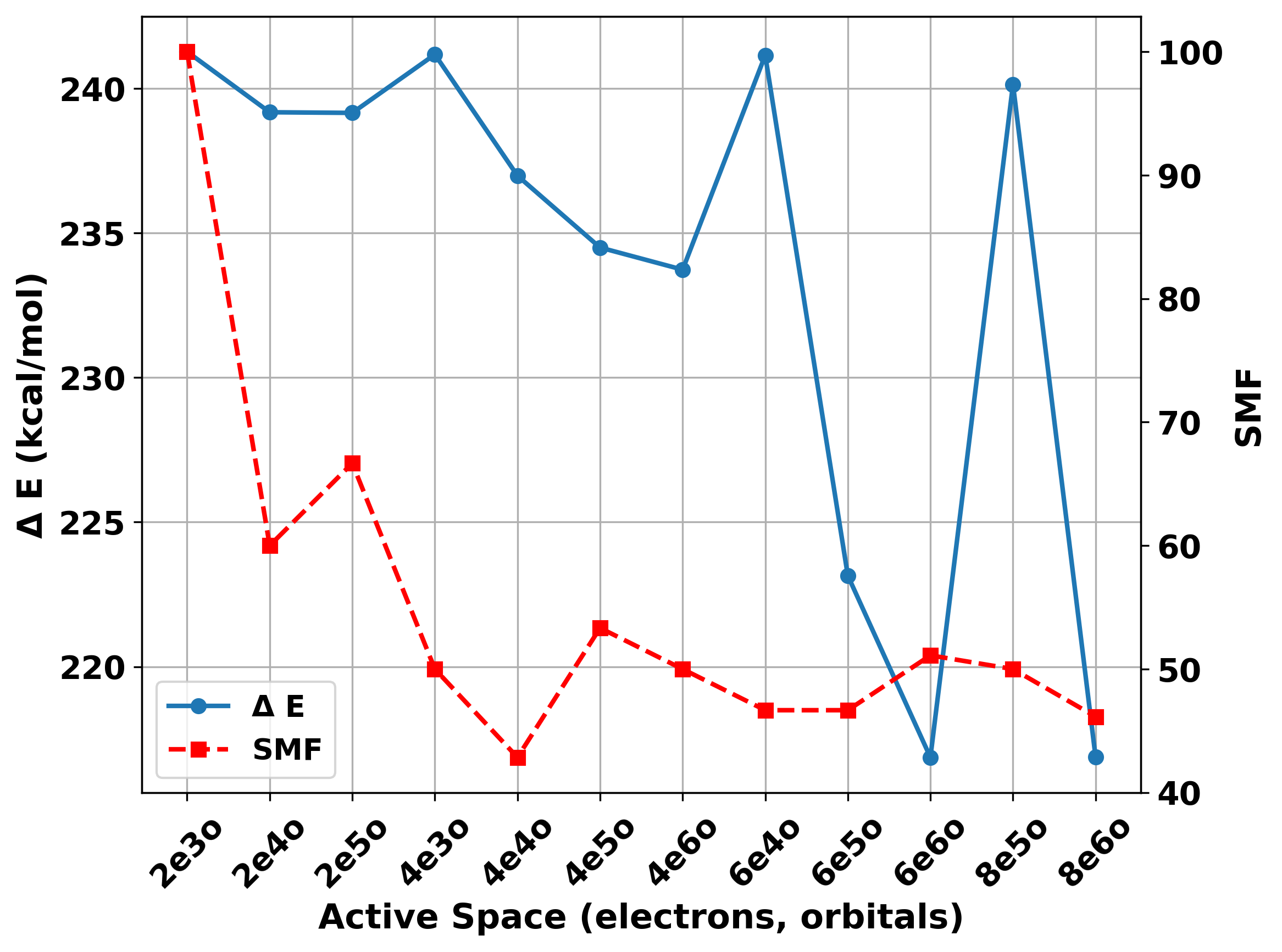}
        \caption{}
        \label{fig:hf_prob}
    \end{subfigure}

    % \vspace{0.3cm}

    % ---------- Row 2 ----------
    \begin{subfigure}{0.4\textwidth}
        \centering
        \includegraphics[width=0.9\linewidth]{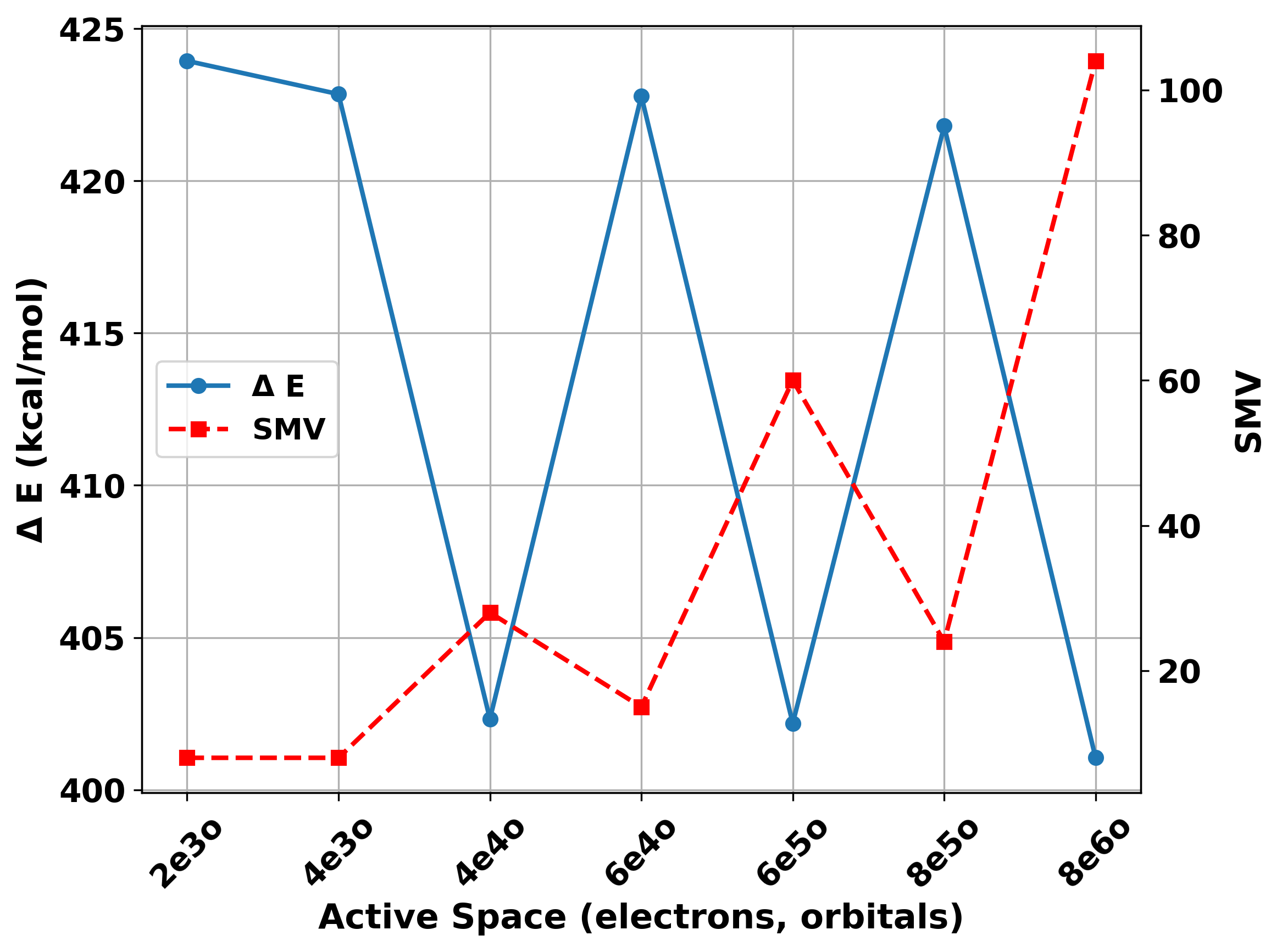}
        \caption{}
        \label{fig:h2_energy}
    \end{subfigure}
%     %% \hfill
    \begin{subfigure}{0.4\textwidth}
        \centering
        \includegraphics[width=0.9\linewidth]
        {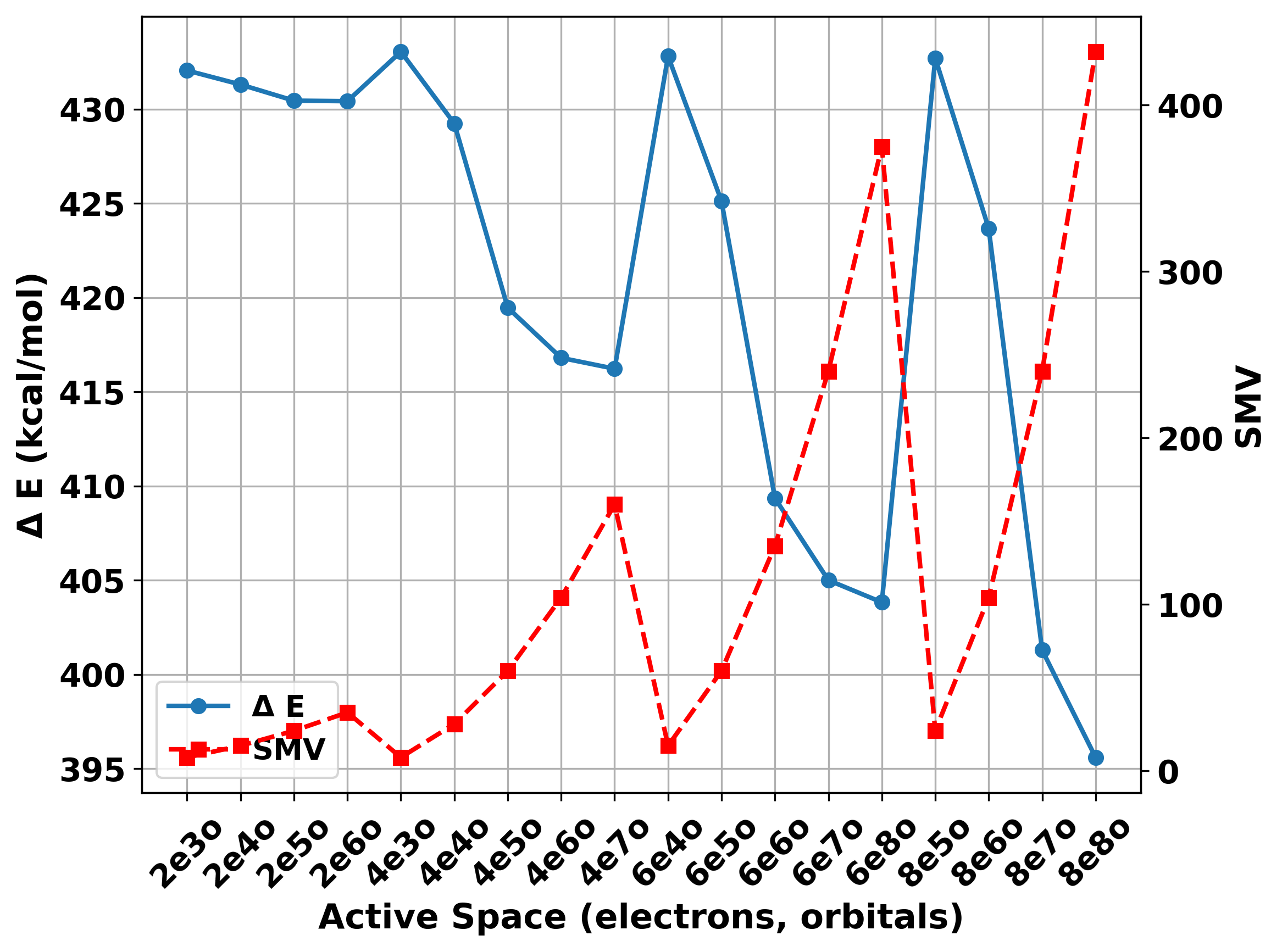}
        \caption{}
        \label{fig:h2_prob}
    \end{subfigure}

    \caption{Absolute Energy Error and SMF/SMV values for (a) ene, (b) eneophile, (c) product, and (d) transition state.}
    \label{fig:main2}
\end{figure}

\begin{table}[]
\caption{\label{comp-ae} Reaction and Activation energies (in kcal/mol) for Alder-ene Reaction and their differences}
\begin{ruledtabular}
\begin{tabular}{lccr}
\textrm{Energy}&
\textrm{CCSD }&
\textrm{SMF-VQE }&
\textrm{Diff.}\\
\colrule
Reaction & -58.29151 & -59.08381 & 0.79230 \\
Activation & 34.30900 & 28.06700 & 6.24199 \\
\end{tabular}
\end{ruledtabular}
\end{table}

\begin{figure}[]
    \centering
    \includegraphics[width=0.9\linewidth]{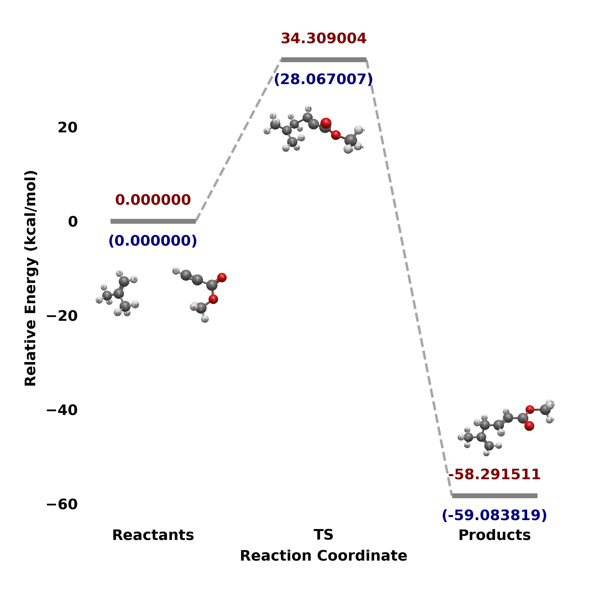}
    \caption{Comparison of PES for CCSD (red) and SMF-VQE (blue and inside bracket) for Alder-ene Reaction}
    \label{pes2}
\end{figure}

\subsection{Alder-Ene Reaction}
We chose the Alder-Ene reaction in Fig.\ref{da-ae}(b) as it has an ester group -COOMe as an electron withdrawing group (EWG) as an eneophile, which makes it electron deficient and highly reactive to the ene, which contains an allylic hydrogen and a C=C bond that donates electron density through hyperconjugation, making the transition state and product more viable and stable. The enophile possesses $C_s$ symmetry and therefore, as per table \ref{variable-ae}, the active spaces are generated from a pool of four occupied and four vacant orbitals, and the active spaces generated will contain a maximum of six orbitals and a minimum of three orbitals, resulting in 12 candidate active spaces. The ene and product possess $C_1$ symmetry and therefore, as per table \ref{variable-ae}, the active spaces are generated from a pool of four occupied and two vacant orbitals, and the active spaces generated will contain a maximum of six orbitals and a minimum of three orbitals, resulting in 7 candidate active spaces for both species. The transition state possesses $C_1$ symmetry and therefore, as per table \ref{variable-ae}, the active spaces are generated from a pool of four occupied and five vacant orbitals, and the active spaces generated will contain a maximum of eight orbitals and a minimum of three orbitals, resulting in 18 candidate active spaces. VQE calculations were performed for all candidate active spaces of the ene, eneophile, product, and transition state species involved in the Alder-Ene reaction. Figure \ref{fig:main2} presents all of these active spaces and the absolute energy errors of all VQE calculations relative to CCSD, together with the corresponding SMF values for the enophile and SMV values for the ene, product, and transition state. Absolute energies are provided in the Supporting Information. Similar to the Diels-Alder reaction, Figures \ref{fig:main2}(a)-(d) show that, for several active-space sequences with fixed electron number, a partial reduction in the absolute energy deviations is observed with increasing orbital number, although the overall behavior remains non-monotonic. The VQE electronic energies exhibit substantial deviations from CCSD, with average absolute deviations of approximately 161, 230, 412, and 415 kcal/mol for the ene, eneophile, product, and transition state, respectively. The absolute energy deviations also exhibit non-monotonic variations across different active spaces. \par

\begin{figure}
     \centering
     \includegraphics[width=0.5\linewidth]{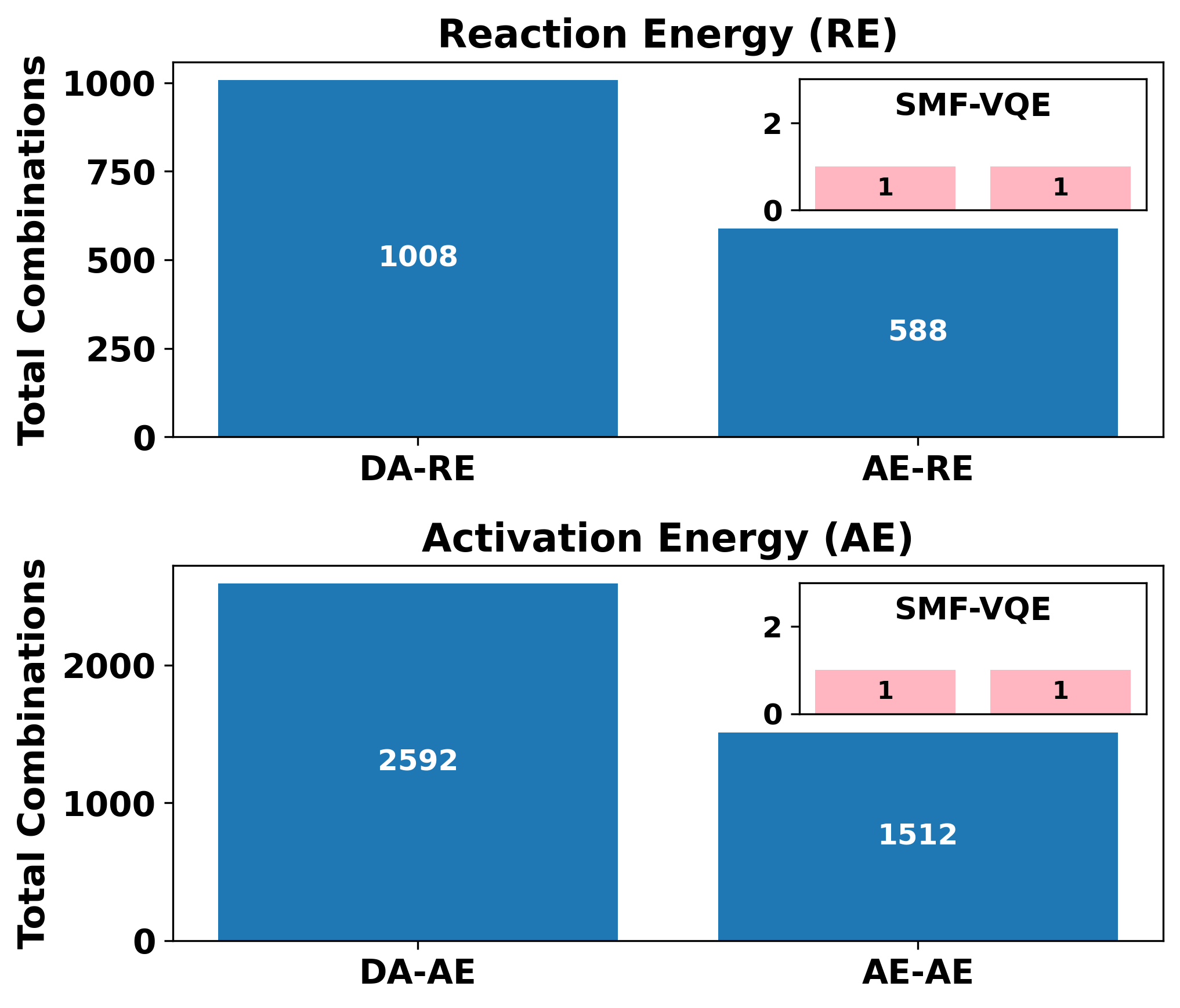}
     \caption{Combinatorial Reduction}
     \label{comb}
 \end{figure}

The reaction and activation energies were calculated similarly, as defined in Equations \ref{eq:rxn} and \ref{eq:actn}, using the maximum SMF/SMV values. The active spaces AS(8e,6o), AS(2e,3o), AS(8e,6o), and AS(8e,8o) were found to possess the largest SMF or SMV values for the ene, eneophile, product, and transition state, respectively. These active spaces were therefore selected for evaluating the reaction and activation energetics within the symmetry-guided framework. A total of 558 possible combinations of reaction energies and 1512 possible combinations of activation energies can be generated from all active-space combinations. The symmetry-guided criterion, however, reduces these possibilities to a single symmetry-consistent combination for each quantity. Figure \ref{comb} illustrates the reduction in the number of possible combinations obtained through the SMF-guided selection. \par
The reaction and activation energies obtained from the selected active spaces are summarized in Table \ref{comp-ae}. Despite the substantial deviations in the absolute ground-state energies, significant cancellation of deviations occurs in the energy differences. The reaction energy differs from the CCSD value by approximately 0.7 kcal/mol, while the activation energy differs by approximately 6 kcal/mol,  which can be seenin Figure \ref{pes2}. These results indicate that symmetry-guided active-space selection can provide chemically meaningful reaction energetics for the Alder-Ene reaction despite significant deviations in the underlying absolute VQE energies.

\subsection{Interpretation of Symmetry-Guided Energetics and Resource Scaling}
The present results reveal several important features regarding the role of symmetry-guided active-space selection in VQE calculations of chemically complex pericyclic reactions. Although the absolute VQE electronic energies exhibit substantial deviations from the corresponding CCSD values for all species, the reaction and activation energetics obtained from the symmetry-guided active spaces remain in substantially improved agreement with the classical reference calculations. This behavior indicates that the energetics obtained from the selected active spaces retain comparable deviations across the reaction coordinate, leading to significant cancellation upon taking energy differences. \par

An important observation emerging from Figures 2(a-d) is that the deviations do not vary monotonically with active-space size. In several cases, enlarging the active space does not necessarily improve the resulting energetics within the truncated VQE framework. This behavior highlights the sensitivity of near-term quantum simulations to the structure of the chosen excitation manifold and indicates that chemically meaningful active spaces cannot be identified solely on the basis of active-space size. Instead, the symmetry properties of the excitation operators play an important role in identifying physically relevant active spaces for reaction-energy calculations. The active spaces selected through the maximum SMF/SMV criterion also exhibit chemically differentiated behavior across the reaction coordinate. For the Diels-Alder reaction, the diene and dienophile  species favor comparatively compact active spaces, AS(4e,4o) and AS(2e,4o), respectively, whereas the product and transition state require substantially larger active spaces, AS(8e,6o) and AS(8e,8o). Similarly, for the Alder-Ene reaction, the enophile favors the compact AS(2e,3o) active space, while the ene and product select AS(8e,6o), and the transition state again selects the largest active space considered in the present work, namely AS(8e,8o). These observations indicate that the symmetry-guided framework adapts non-uniformly across the reaction coordinate and does not simply favor the largest active spaces for all molecular species. \par

\begin{table}
\caption{\label{resource-table} Selected active spaces and variational resource scaling for the species involved in the Diels-Alder and Alder-Ene reactions.}
\begin{ruledtabular}
\begin{tabular}{lcccc}
Species & Selected AS & Criterion & Sym. Excitations & Tapered Params \\
\colrule
Diene & 4e,4o & SMF = 57.14 & 16 & 14 \\
dienophile  & 2e,4o & SMF = 66.67 & 8 & 15 \\
Product & 8e,6o & SMV = 104 & 104 & 92 \\
TS & 8e,8o & SMV = 432 & 432 & 360 \\
\colrule
Ene & 8e,6o & SMV = 104 & 104 & 92 \\
enophile & 2e,3o & SMF = 100 & 8 & 4 \\
Product & 8e,6o & SMV = 104 & 104 & 92 \\
TS & 8e,8o & SMV = 432 & 432 & 360 \\
\end{tabular}
\end{ruledtabular}
\end{table}

To further examine the variational complexity associated with the selected active spaces, the total excitation operators, symmetry-matched excitations, tapered mapped qubits, and tapered mapped variational parameters corresponding to the selected active spaces are summarized in Table~V.  For the $C_{2v}$ and $C_s$ species, the selected active spaces maximize the SMF values, whereas for the $C_1$ species all excitations belong to the same irreducible representation and therefore SMV is used to distinguish the active spaces. The results show a substantial increase in the excitation manifolds and variational parameter spaces along the reaction coordinate, particularly near the transition-state region. In particular, the transition states of both reactions involve 432 symmetry-matched excitations and 360 tapered mapped variational parameters, whereas the dienophile and enophile species involve only 8 symmetry-matched excitations with 15 and 4 tapered mapped parameters, respectively, as parity mapper prunes qubits and tapers off those parameters which become redundant after the pruning of qubits and this pruning and tapering varies from one molecule to another and also with respect to their active spaces. The products and the ene species occupy an intermediate regime with 104 symmetry-matched excitations, and 92 tapered mapped parameters. An interesting observation is that the ene species already exhibits excitation and parameter counts comparable to those of the products, unlike the comparatively compact dienophile and enophile species. This behavior suggests that the electronic structure of the ene species involves a more distributed excitation manifold even at the reactant stage, consistent with the extended conjugated character associated with the Alder-Ene framework. The substantially larger excitation manifolds selected for the transition states are also chemically consistent with the concerted nature of pericyclic reactions. Simultaneous bond formation and bond breaking near the transition-state region lead to increased orbital mixing and near-degeneracy effects, thereby requiring broader symmetry-consistent excitation sectors within the variational description. In contrast, simpler reactant species such as the dienophile and enophile retain comparatively localized electronic structures and therefore involve substantially smaller excitation manifolds. The ene species represents an intermediate situation in which extended conjugation and allylic participation already lead to excitation and parameter counts comparable to those of the products. These results quantitatively demonstrate the rapid growth in variational complexity associated with electronically more demanding regions of the reaction coordinate. The transition states, therefore, represent the most resource-intensive species within the present study, requiring the largest active spaces, excitation manifolds, and variational parameter sectors. \par
The larger deviations observed for activation energies compared to the corresponding reaction energies are also consistent with this behavior. Transition states involve simultaneous bond formation and bond breaking and are therefore expected to be more sensitive to active-space truncation and correlation treatment. The large excitation manifolds selected for the transition states further reflect the increased electronic complexity associated with these regions of the potential energy surface. Overall, the present analysis indicates that symmetry-guided active-space selection provides a physically meaningful framework for identifying chemically relevant excitation manifolds in resource-constrained VQE calculations. Even though the absolute electronic energies remain substantially shifted relative to CCSD, the selected symmetry-guided active spaces retain chemically meaningful relative energetics for both the Diels-Alder and Alder-Ene reactions and their corresponding transition states.

\section{Conclusion}
In this work, we investigated Diels-Alder and Alder-Ene reactions as benchmark systems for assessing symmetry-guided quantum simulations of chemically more complex pericyclic reactions involving larger active spaces, extended $\pi$ interactions, and transition states. Using VQE calculations within a symmetry-guided active-space framework, reaction and activation energetics were evaluated for six molecular species and two transition states under present quantum resource constraints. \par
The results show that although the absolute VQE electronic energies exhibit substantial deviations from CCSD, symmetry-guided active-space selection yields significantly improved agreement for reaction and activation energetics. In particular, reaction energies were reproduced at or near chemical accuracy, while activation energies remained within approximately $5-6$ kcal/mol of the CCSD reference values. The larger deviations observed for transition states are consistent with their near-degeneracy and partial bond-breaking character, which require larger and more delicate active-space descriptions. Transition states are also expected to be more sensitive to active-space truncation because of increased multi-configurational character associated with simultaneous bond formation and bond breaking. The present study also demonstrates that symmetry-guided selection substantially reduces the combinatorial choices of active spaces across the reaction coordinate. For the Diels-Alder and Alder-Ene reactions, thousands of possible combinations of reaction and activation energetics were reduced to a single symmetry-consistent combination for each case. These results suggest that chemically complex pericyclic reactions provide meaningful benchmark systems for assessing symmetry-guided quantum simulations within present computational and quantum resource limitations. \par
Future work will examine the robustness of symmetry-guided active-space selection in larger basis sets and under realistic hardware noise conditions, including the role of error-mitigation strategies on quantum devices.

\begin{acknowledgments}
MS and AK acknowledge support from the Indian Institute of Technology, Jodhpur, to provide the facilities necessary to complete the work. MS acknowledges the Department of Chemistry, IIT Jodhpur, HPC Services, IIT Jodhpur, and MoE for providing research facilities and financial support. 
\end{acknowledgments}

% \newpage
\bibliography{main}

\end{document}

\setcounter{page}{1}
\renewcommand{\thepage}{S\arabic{page}}
\setcounter{figure}{0}   % Reset the figure counter to start from S1
\renewcommand{\thefigure}{S\arabic{figure}}
\renewcommand{\figurename}{Figure}
\setcounter{table}{0}
\renewcommand{\thetable}{S\arabic{table}}
\renewcommand{\tablename}{Table}

% \preprint{APS/123-QED}

\title{\Huge\bfseries Computing Reaction and Activation Energies of Pericyclic Reactions using a Symmetry-Adapted VQE Algorithm : Supporting Information}% Force line breaks with \\

\author{\bfseries Maitreyee Sarkar$^{1}$}
% \altaffiliation[Also at ]{Physics Department, XYZ University.}
\email{sarkar.5@iitj.ac.in}
\author{\bfseries Manikandan Paranjothy$^{2}$}%
\email{pmanikandan@iitj.ac.in}
\author{\bfseries Atul Kumar$^{1}$}%
\email{atulk@iitj.ac.in}
\affiliation{%
$^{1}$ Quantum Information and Computation Lab, Department of Chemistry, Indian Institute of Technology Jodhpur, Rajasthan, India, 342030 %\textbackslash\textbackslash
}%
\affiliation{%
$^{2}$ Chemical Dynamics Research Group, Department of Chemistry, Indian Institute of Technology Jodhpur, Rajasthan, India, 342030 %\textbackslash\textbackslash
}
\maketitle

\section{Optimized Geometries}
Geometry optimization values for all reactants and products, calculated using DFT/xc m05-2x and cc-pVDZ basis set with NWChem, where the coordinates are expressed in Angstrom units.

\subsection{Diels-Alder Reaction}

\begin{figure}[h]
\centering
\begin{subfigure}{0.3\textwidth}
\centering
\includegraphics[width=4cm]{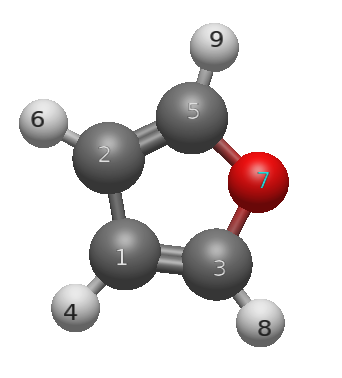} 
% \caption{\ce{Cl2} molecule}
\end{subfigure}
\hspace{0.05\textwidth} % Add some horizontal space between the subfigures
\begin{subfigure}{0.3\textwidth}
\centering
\includegraphics[height=4cm]{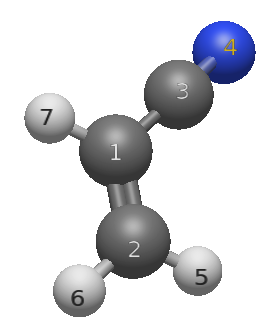}
% \caption{\ce{HF} molecule}
\end{subfigure}
\caption{Diene and Dieneophile}
\end{figure}

\begin{table}[h]
\begin{tabular}{ccccccccccc}
   C  &   0.00000000  &  -0.71587689  &  -0.95220267 & &&& C  &  -0.52781233  &  -0.69422379  &  -0.18931269 \\
   C  &   0.00000000  &   0.71587689  &  -0.95220267 & &&& C  &   0.21616115  &  -1.68711762  &   0.27226713 \\
   C  &   0.00000000  &  -1.08905462  &   0.34595789 & &&& C  &  -0.21481500  &   0.68096621  &   0.07705371 \\
   H  &   0.00000000  &  -1.37549839  &  -1.80630141 & &&& N  &   0.02020207  &   1.78269958  &   0.27959801 \\
   C  &   0.00000000  &   1.08905462  &   0.34595789 & &&& H  &   1.09826639  &  -1.49962348  &   0.87482500 \\
   H  &   0.00000000  &   1.37549839  &  -1.80630141 & &&& H  &  -0.04485648  &  -2.71660207  &   0.05455591 \\
   O  &   0.00000000  &   0.00000000  &   1.14952390 & &&& H  &  -1.41178794  &  -0.87096476  &  -0.79268533 \\
   H  &   0.00000000  &  -2.04828382  &   0.83977373 \\
   H  &   0.00000000  &   2.04828382  &   0.83977373 \\
\end{tabular}
\end{table}

\begin{figure}[h]
\centering
\begin{subfigure}{0.3\textwidth}
\centering
\includegraphics[width=4cm]{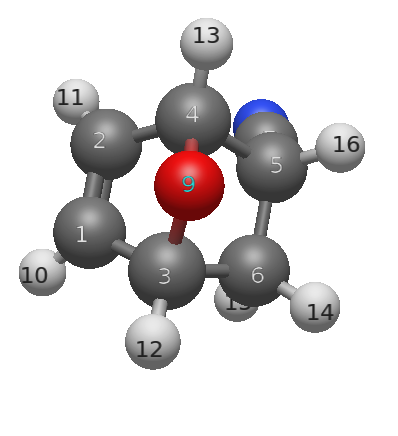} 
% \caption{\ce{Cl2} molecule}
\end{subfigure}
\hspace{0.05\textwidth} % Add some horizontal space between the subfigures
\begin{subfigure}{0.3\textwidth}
\centering
\includegraphics[height=4cm]{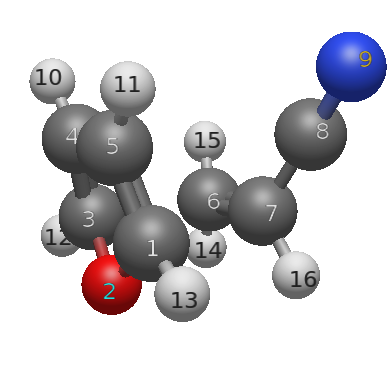}
% \caption{\ce{HF} molecule}
\end{subfigure}
\caption{Product and TS}
\end{figure}

\begin{table}[h]
\begin{tabular}{ccccccccccc}
   C  &  -0.68053354  &  -1.62990280  &  -0.82908655 & &&& C  &  -0.48629522  &  -0.69078804  &   1.10091304 \\
   C  &  -1.43669139  &  -0.83217859  &  -0.08661810 & &&& O  &   0.56182746  &  -1.50889083  &   0.83655207 \\
   C  &   0.70136045  &  -1.57523851  &  -0.20618268 & &&& C  &   0.61501488  &  -1.53918974  &  -0.53308205 \\
   C  &  -0.51121261  &  -0.30340526  &   0.99247347 & &&& C  &  -0.69244045  &  -1.28490803  &  -1.03087582 \\
   C  &   0.44133101  &   0.74459066  &   0.30362459 & &&& C  &  -1.38950871  &  -0.72217234  &   0.01337094 \\
   C  &   1.30586572  &  -0.19194363  &  -0.58731856 & &&& C  &   1.40909371  &   0.32464923  &  -0.66330465 \\
   C  &  -0.26999720  &   1.78911660  &  -0.42085683 & &&& C  &   0.74167169  &   1.00723108  &   0.37070528 \\
   N  &  -0.84418744  &   2.60417697  &  -0.98746811 & &&& C  &  -0.24344165  &   2.00003533  &   0.05508970 \\
   O  &   0.40092672  &  -1.37203968  &   1.17438205 & &&& N  &  -1.05889386  &   2.78239375  &  -0.20246757 \\
   H  &  -0.93242458  &  -2.12799409  &  -1.75265353 & &&& H  &  -1.00185867  &  -1.37715350  &  -2.06580008 \\
   H  &  -2.45386385  &  -0.51397391  &  -0.25195296 & &&& H  &  -2.36893596  &  -0.25925565  &  -0.00662775 \\
   H  &   1.35743919  &  -2.42755058  &  -0.34050429 & &&& H  &   1.35913661  &  -2.21078148  &  -0.94848427 \\
   H  &  -0.94219409  &   0.02291146  &   1.93170751 & &&& H  &  -0.68339254  &  -0.49094650  &   2.14832101 \\
   H  &   2.35045689  &  -0.13623603  &  -0.29179962 & &&& H  &   2.43096727  &   0.00521330  &  -0.46872800 \\
   H  &   1.21950144  &   0.03103308  &  -1.64720307 & &&& H  &   1.19895294  &   0.60075171  &  -1.69386040 \\
   H  &   1.05445240  &   1.21406635  &   1.07022311 & &&& H  &   1.23245805  &   1.13422341  &   1.33248387 \\
\end{tabular}
\end{table}

\FloatBarrier
% \newpage

\subsection{Alder-ene Reaction}

\begin{figure}[h]
\centering
\begin{subfigure}{0.3\textwidth}
\centering
\includegraphics[width=4cm]{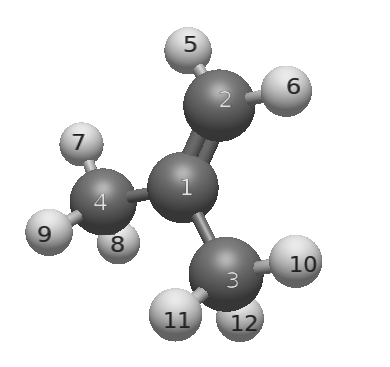} 
% \caption{\ce{Cl2} molecule}
\end{subfigure}
\hspace{0.05\textwidth} % Add some horizontal space between the subfigures
\begin{subfigure}{0.3\textwidth}
\centering
\includegraphics[height=4cm]{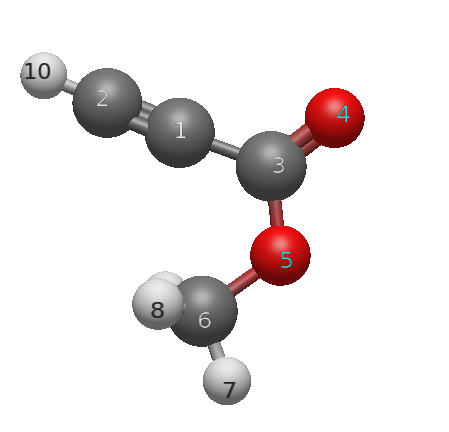}
% \caption{\ce{HF} molecule}
\end{subfigure}
\caption{Ene and Eneophile}
\end{figure}

\begin{table}[h]
\begin{tabular}{ccccccccccc}
   C  &   0.13643224  &  -0.03343709  &  -0.06499054 &&&& C  &  -1.16133660  &  -0.20642820  &   0.00000000 \\
   C  &   1.45663148  &  -0.10182188  &  -0.15527526 &&&& C  &  -2.12585065  &  -0.91019359  &   0.00000000 \\
   C  &  -0.60504688  &   1.27357475  &  -0.10178166 &&&& C  &  -0.01212369  &   0.70396762  &   0.00000000 \\
   C  &  -0.72363379  &  -1.25750755  &   0.08115954 &&&& O  &  -0.13951289  &   1.89060545  &   0.00000000 \\
   H  &   1.97417383  &  -1.05616668  &  -0.12669444 &&&& O  &   1.19360581  &   0.12484804  &   0.00000000 \\
   H  &   2.06081840  &   0.79446750  &  -0.26048018 &&&& C  &   1.29988857  &  -1.30369231  &   0.00000000 \\
   H  &  -0.12596905  &  -2.17003413  &   0.10077484 &&&& H  &   2.36598449  &  -1.51848170  &   0.00000000 \\
   H  &  -1.31154719  &  -1.20394055  &   1.00342242 &&&& H  &   0.83327686  &  -1.72372395  &   0.89284082 \\
   H  &  -1.43764399  &  -1.32420759  &  -0.74640206 &&&& H  &   0.83327686  &  -1.72372395  &  -0.89284082 \\
   H  &   0.07505951  &   2.11994890  &  -0.20853994 &&&& H  &  -2.99726444  &  -1.52613113  &   0.00000000 \\
   H  &  -1.31479260  &   1.28721174  &  -0.93556505 \\
   H  &  -1.18976166  &   1.40745742  &   0.81433978 \\
\end{tabular}
\end{table}

\begin{figure}[h]
\centering
\begin{subfigure}{0.3\textwidth}
\centering
\includegraphics[width=6cm]{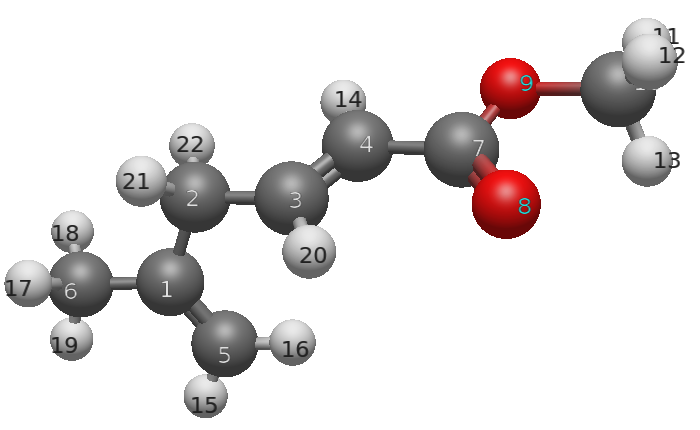} 
% \caption{\ce{Cl2} molecule}
\end{subfigure}
\hspace{0.05\textwidth} % Add some horizontal space between the subfigures
\begin{subfigure}{0.3\textwidth}
\centering
\includegraphics[height=4cm]{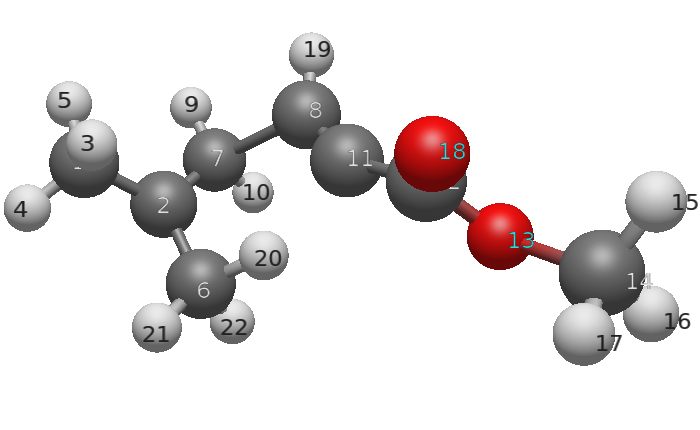}
% \caption{\ce{HF} molecule}
\end{subfigure}
\caption{Product and TS}
\end{figure}

\begin{table}[h]
\begin{tabular}{ccccccccccc}
   C  &   0.35624905  &   2.86293792  &  -0.17578301 &&&& C  &  -0.86573935  &   3.30604197  &   0.67461895 \\
   C  &  -0.57175334  &   1.88898734  &   0.57296440 &&&& C  &  -0.16462333  &   2.53703006  &  -0.40567772 \\
   C  &  -0.57082128  &   0.47997906  &  -0.00215544 &&&& H  &  -1.40786303  &   2.61310635  &   1.34603978 \\
   C  &  -0.16500347  &  -0.60763272  &   0.65199611 &&&& H  &  -1.62209904  &   3.98526041  &   0.24767208 \\
   C  &   1.22790888  &   2.47109953  &  -1.09754207 &&&& H  &  -0.16904362  &   3.89881208  &   1.28443332 \\
   C  &   0.20469010  &   4.32415157  &   0.24346279 &&&& C  &  -0.93217516  &   1.62972232  &  -1.23441309 \\
   C  &  -0.20565454  &  -1.97271878  &  -0.00666239 &&&& C  &   1.21330692  &   2.31653366  &  -0.35991432 \\
   O  &  -0.58282861  &  -2.23531412  &  -1.15837064 &&&& C  &   1.34422465  &   0.60694215  &   0.53998148 \\
   O  &   0.26193253  &  -2.95315371  &   0.90009635 &&&& H  &   1.82385357  &   2.96863893  &   0.27132229 \\
   C  &   0.22054721  &  -4.26767453  &   0.26661434 &&&& H  &   1.71616663  &   1.98166953  &  -1.27139479 \\
   H  &   0.58976633  &  -4.98744190  &   1.01267860 &&&& C  &   0.52286056  &  -0.31571407  &   0.28795201 \\
   H  &  -0.80574227  &  -4.54220779  &  -0.03298501 &&&& C  &  -0.05978445  &  -1.62419021  &   0.43535523 \\
   H  &   0.86186753  &  -4.30678607  &  -0.63095039 &&&& O  &   0.33343357  &  -2.45142942  &  -0.57379571 \\
   H  &   0.20833805  &  -0.57015719  &   1.67923437 &&&& C  &  -0.23584049  &  -3.76665313  &  -0.52484624 \\
   H  &   1.87897095  &   3.18386096  &  -1.60611147 &&&& H  &   0.02016503  &  -4.27511014  &   0.41699004 \\
   H  &   1.33339695  &   1.42360193  &  -1.38368789 &&&& H  &   0.19287432  &  -4.30456660  &  -1.37947588 \\
   H  &  -0.81491910  &   4.67717585  &   0.04128337 &&&& H  &  -1.33306374  &  -3.72448737  &  -0.60940042 \\
   H  &   0.38645930  &   4.43373511  &   1.32048574 &&&& O  &  -0.83018278  &  -1.95942095  &   1.31432597 \\
   H  &   0.91123679  &   4.96006248  &  -0.29921231 &&&& H  &   2.22767359  &   0.79983582  &   1.13841569 \\
   H  &  -0.93275212  &   0.36887021  &  -1.03202316 &&&& H  &  -0.69238740  &   0.60604994  &  -0.69752004 \\
   H  &  -1.59647016  &   2.29371256  &   0.53628899 &&&& H  &  -2.01814235  &   1.78612032  &  -1.24867236 \\
   H  &  -0.27685615  &   1.85632949  &   1.63256667 &&&& H  &  -0.53211351  &   1.46767605  &  -2.24486272 \\
\end{tabular}
\end{table} 

\FloatBarrier
\newpage

\section{Absolute Energy Comparison for Both Reactions}

% new fig
\begin{figure*}[h]
    \centering

    % ---------- Row 1 ----------
    \begin{subfigure}{0.45\textwidth}
        \centering
        \includegraphics[width=\linewidth]{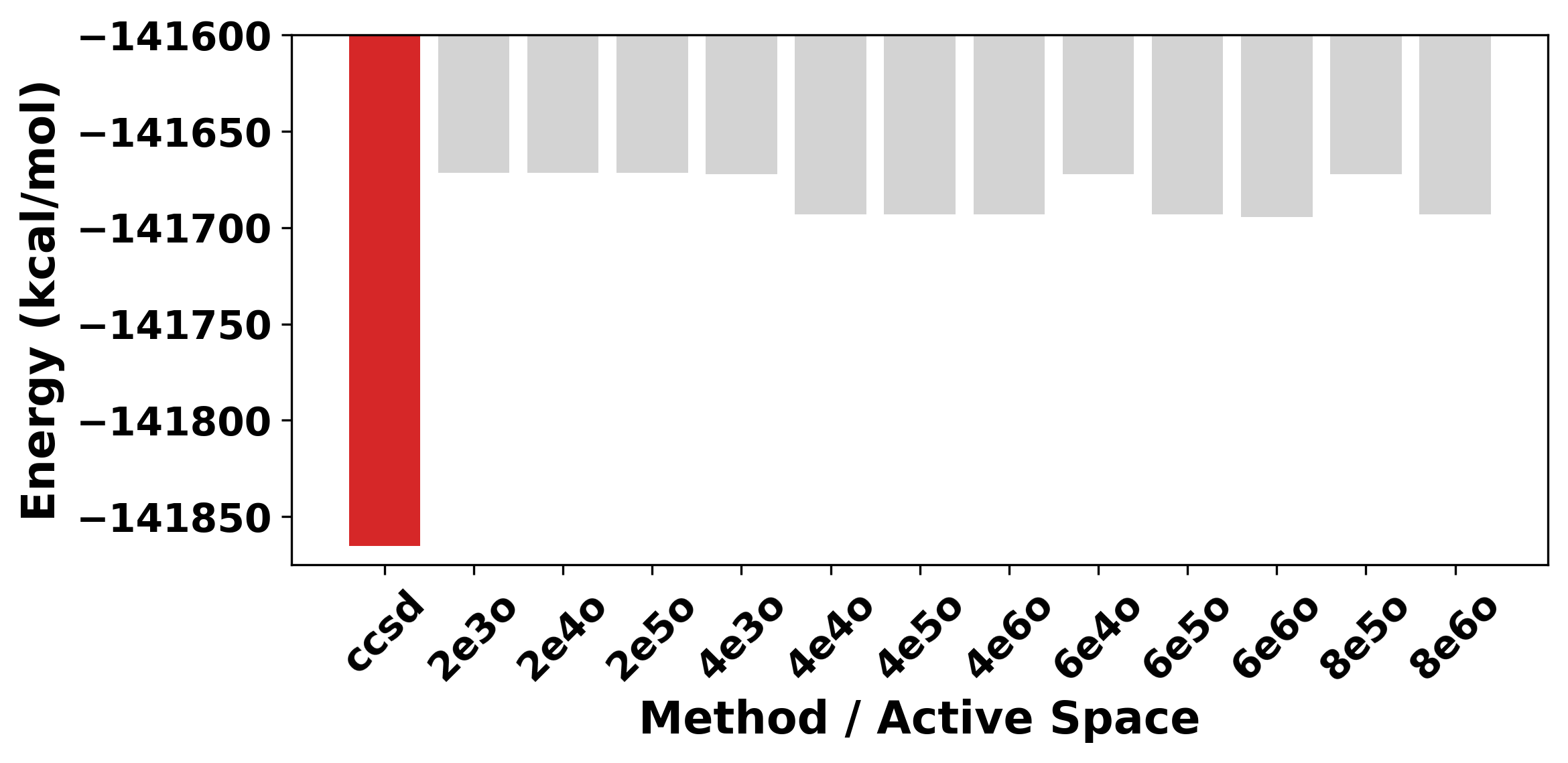}
        \caption{}
        \label{fig:hf_energy}
    \end{subfigure}
    \hfill
    \begin{subfigure}{0.45\textwidth}
        \centering
        \includegraphics[width=\linewidth]{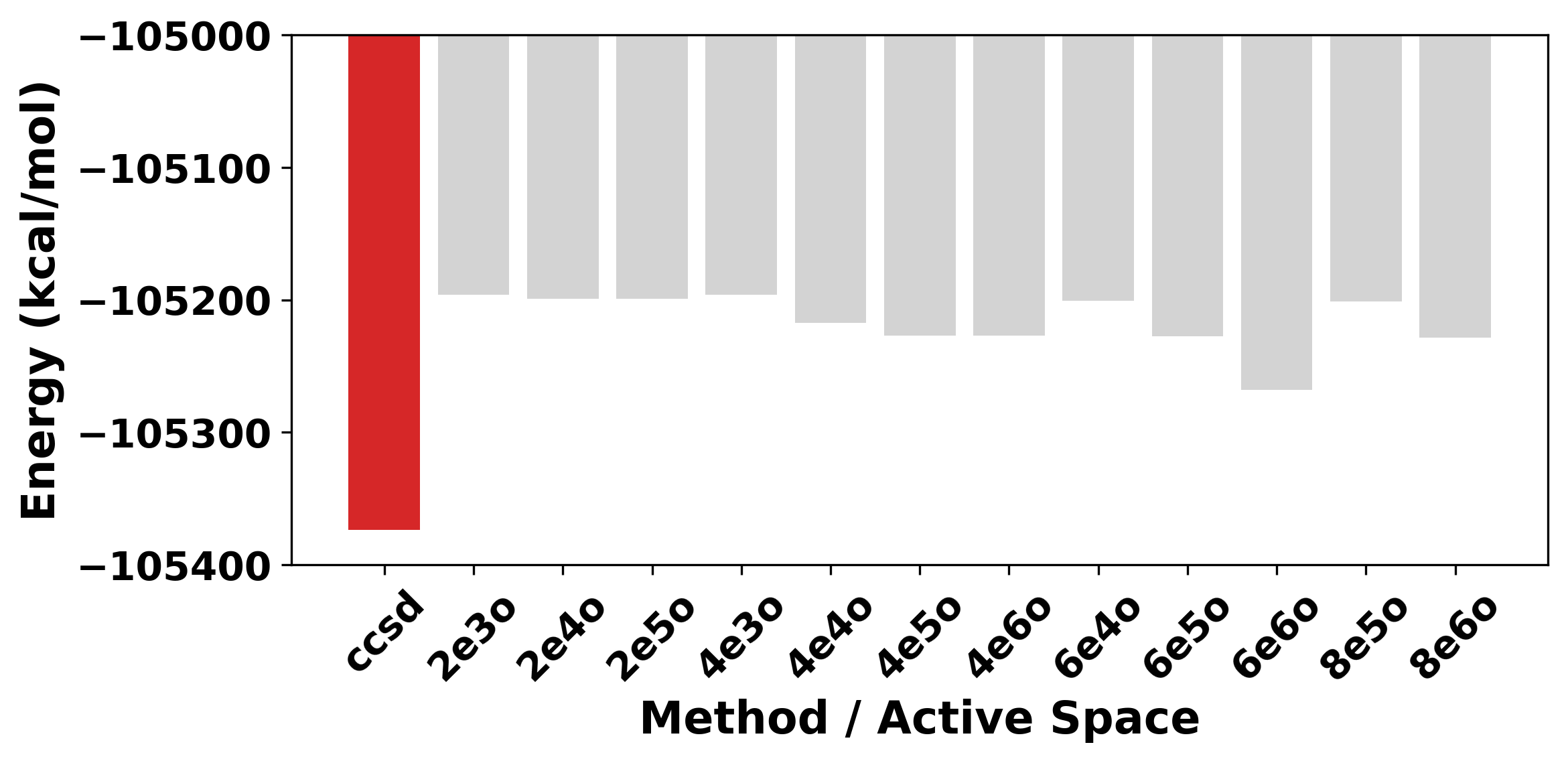}
        \caption{}
        \label{fig:hf_prob}
    \end{subfigure}

    % \medskip

    % ---------- Row 2 ----------
    \begin{subfigure}{0.45\textwidth}
        \centering
        \includegraphics[width=\linewidth]{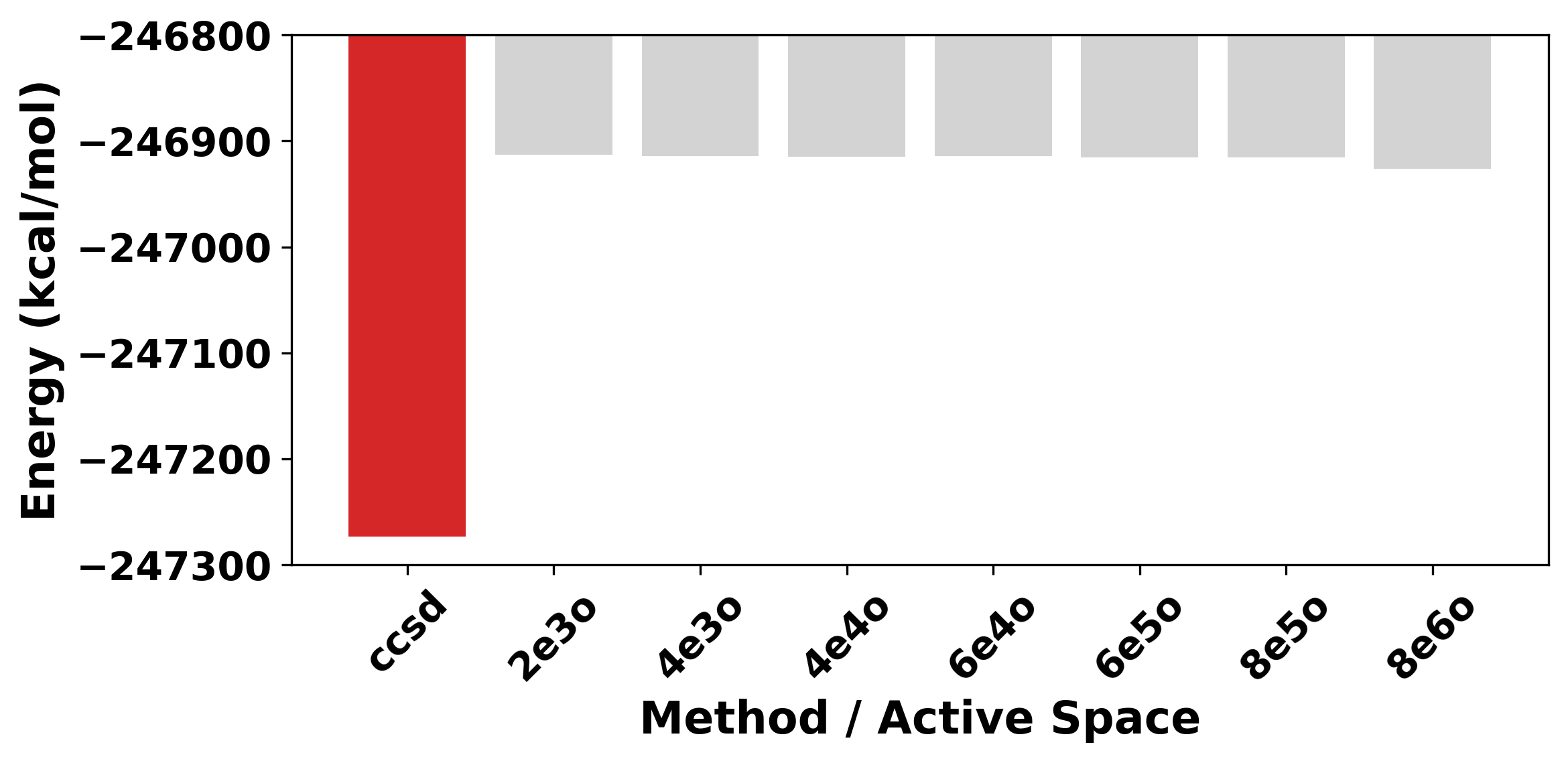}
        \caption{}
        \label{fig:h2_energy}
    \end{subfigure}
    \hfill
    \begin{subfigure}{0.45\textwidth}
        \centering
        \includegraphics[width=\linewidth]{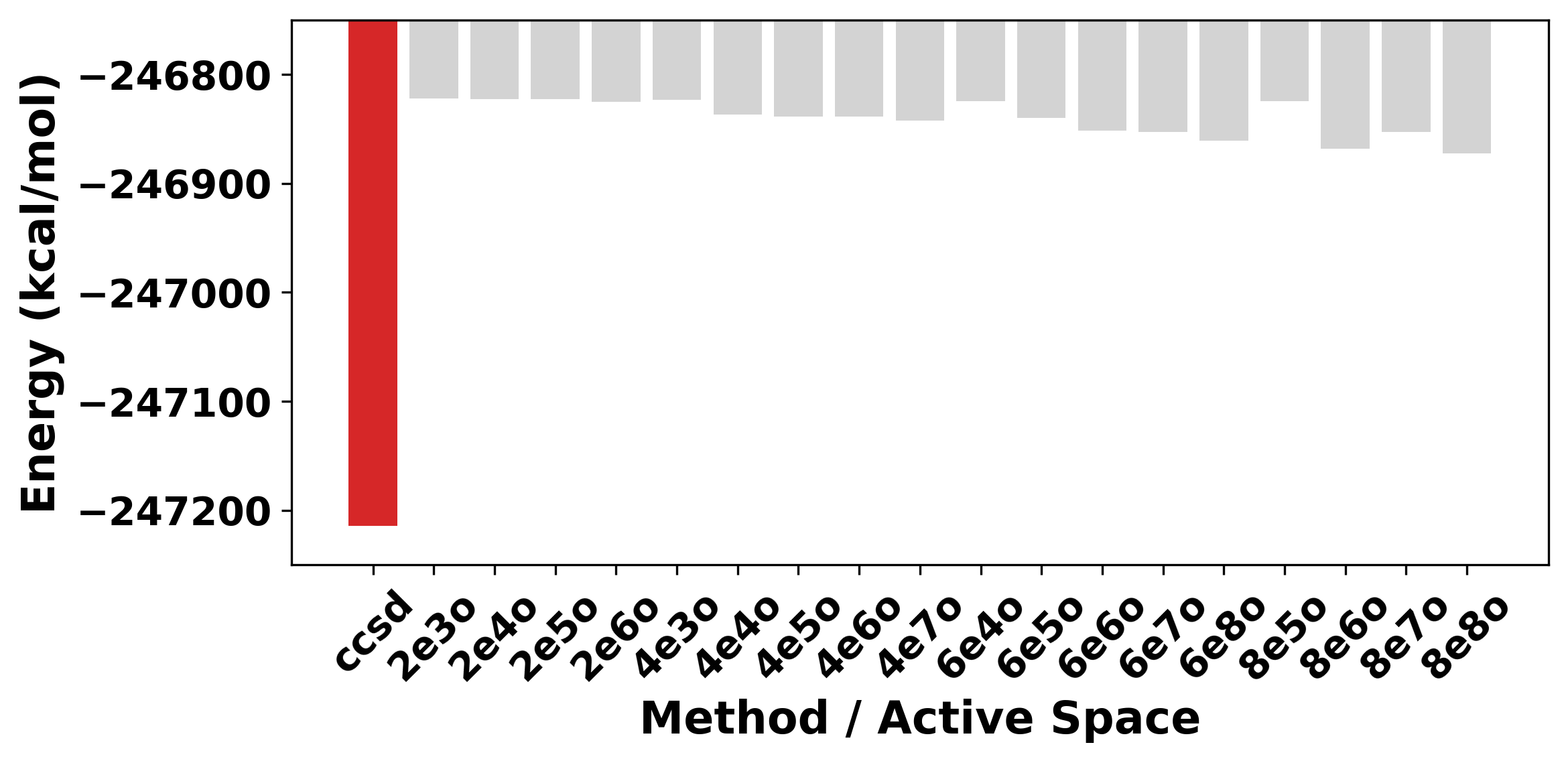}
        \caption{}
        \label{fig:h2_prob}
    \end{subfigure}

    \caption{Comparison of energy values for Diels-Alder Reaction a) diene, b) dieneophile, c) product, and d) transition state.}
    \label{fig:main}
\end{figure*}

% \begin{figure*}
%     \centering
%     \includegraphics[width=0.5\linewidth]{pes_energy.png}
%     \caption{Comparison of PES for CCSD and SMF-VQE for Diels Alder Reaction}
%     \label{fig:placeholder}
% \end{figure*}

% %new fig
\begin{figure*}[h]
    \centering

    % ---------- Row 1 ----------
    \begin{subfigure}{0.45\textwidth}
        \centering
        \includegraphics[width=\linewidth]{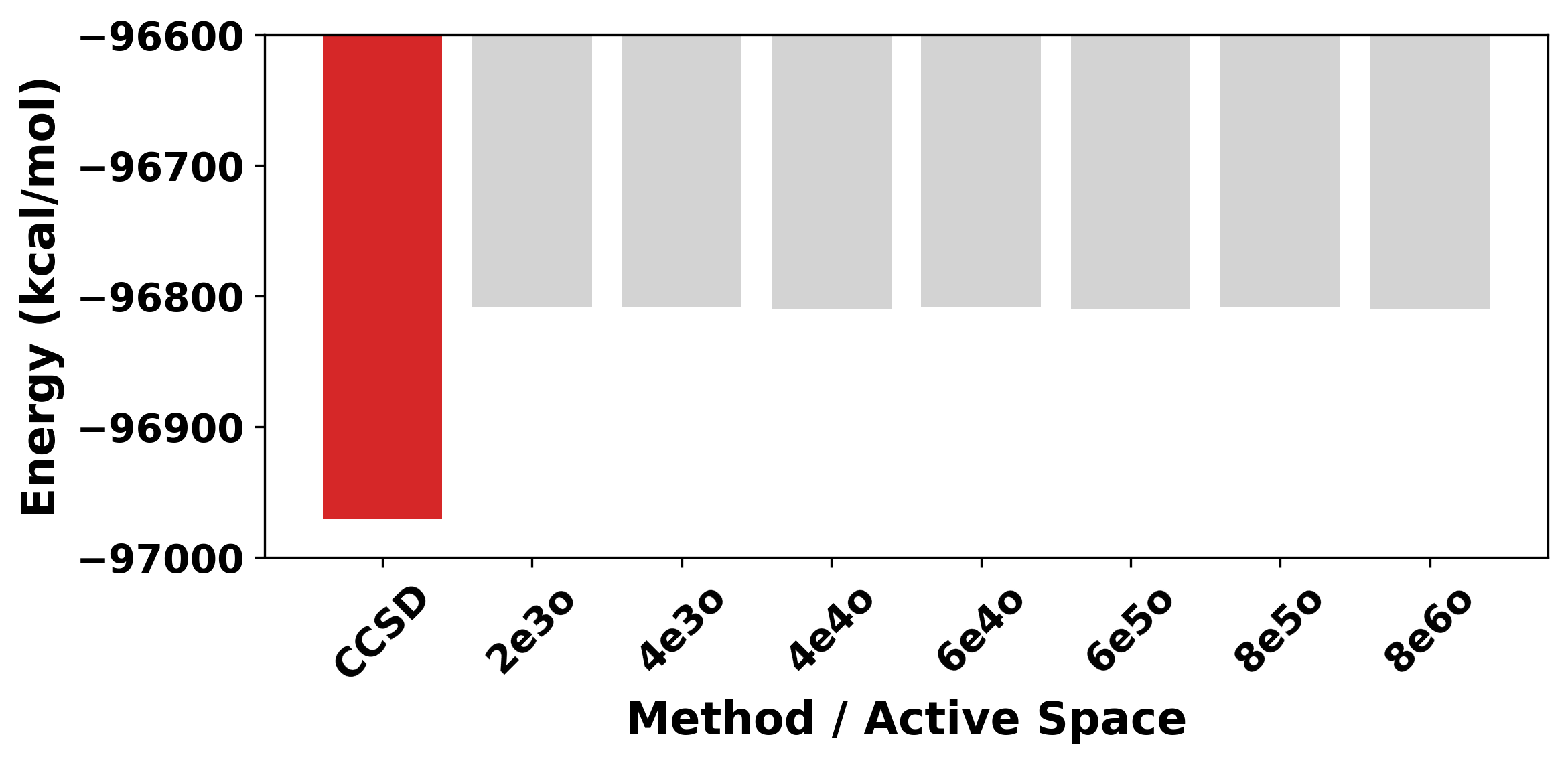}
        % \label{fig:hf_energy}
        \caption{}
    \end{subfigure}
    \hfill
    \begin{subfigure}{0.45\textwidth}
        \centering
        \includegraphics[width=\linewidth]{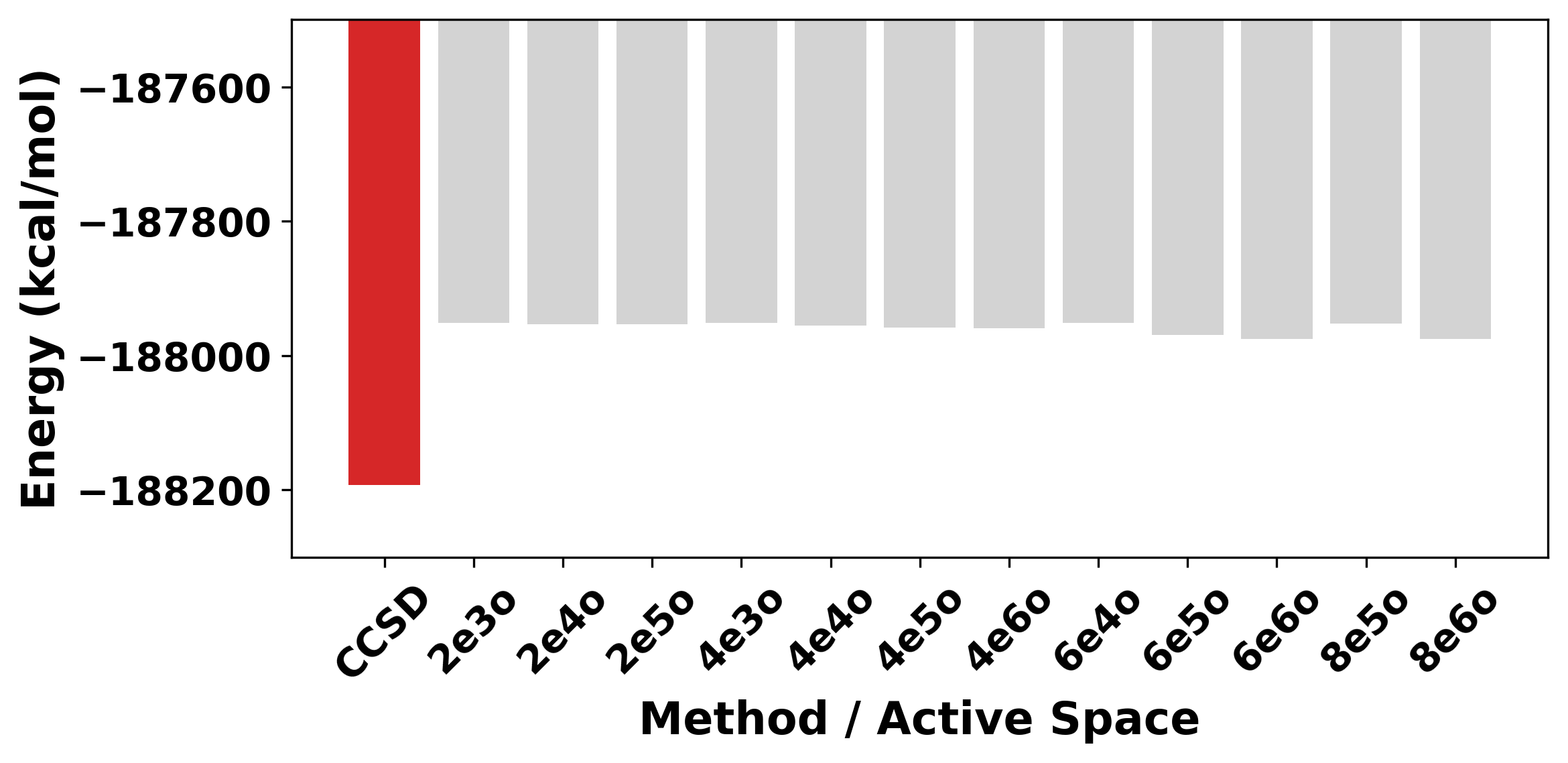}
        % \label{fig:hf_prob}
        \caption{}
    \end{subfigure}

    % \vspace{0.3cm}

    % ---------- Row 2 ----------
    \begin{subfigure}{0.45\textwidth}
        \centering
        \includegraphics[width=\linewidth]{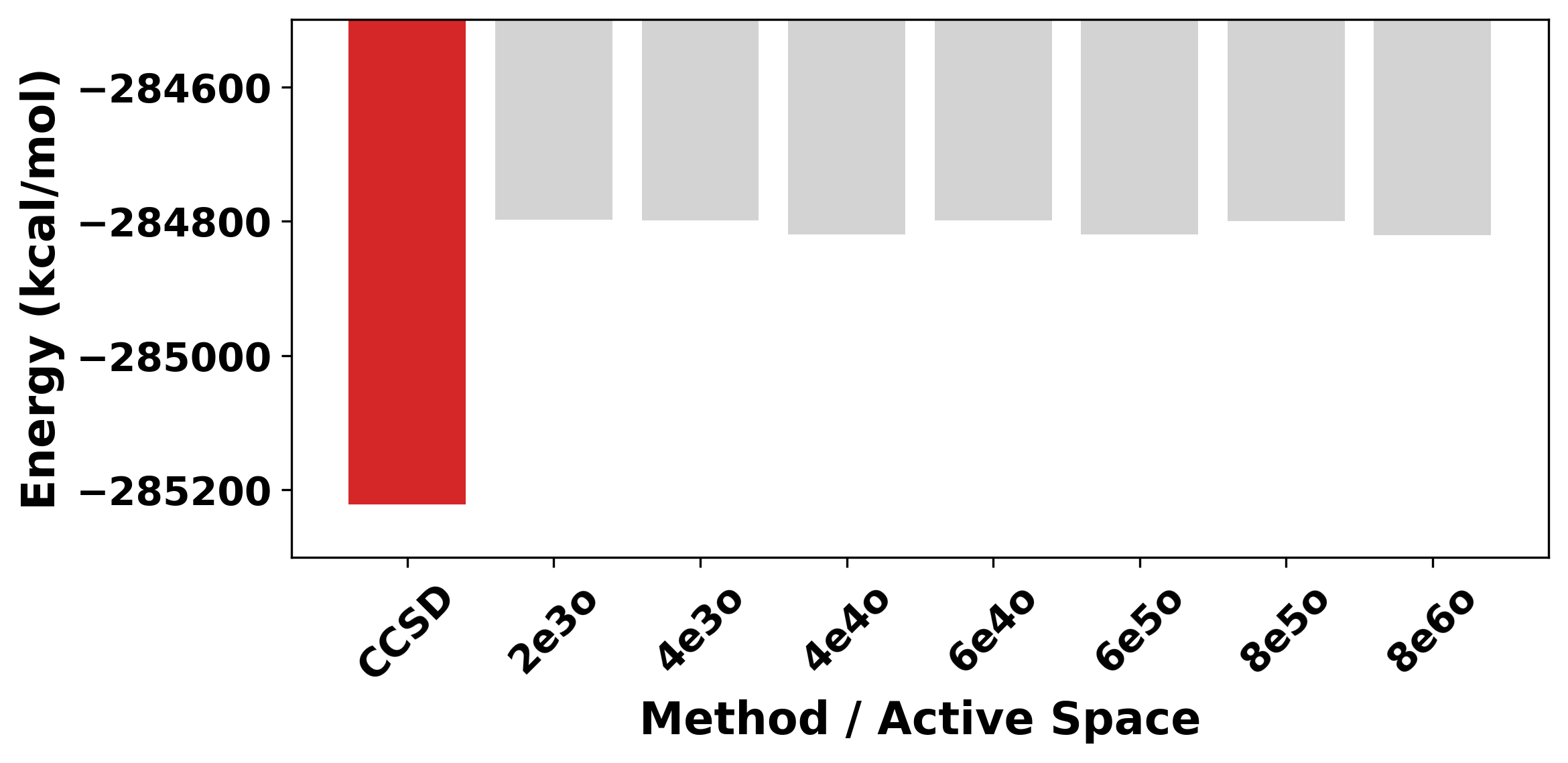}
        % \label{fig:h2_energy}
        \caption{}
    \end{subfigure}
    \hfill
    \begin{subfigure}{0.45\textwidth}
        \centering
        \includegraphics[width=\linewidth]{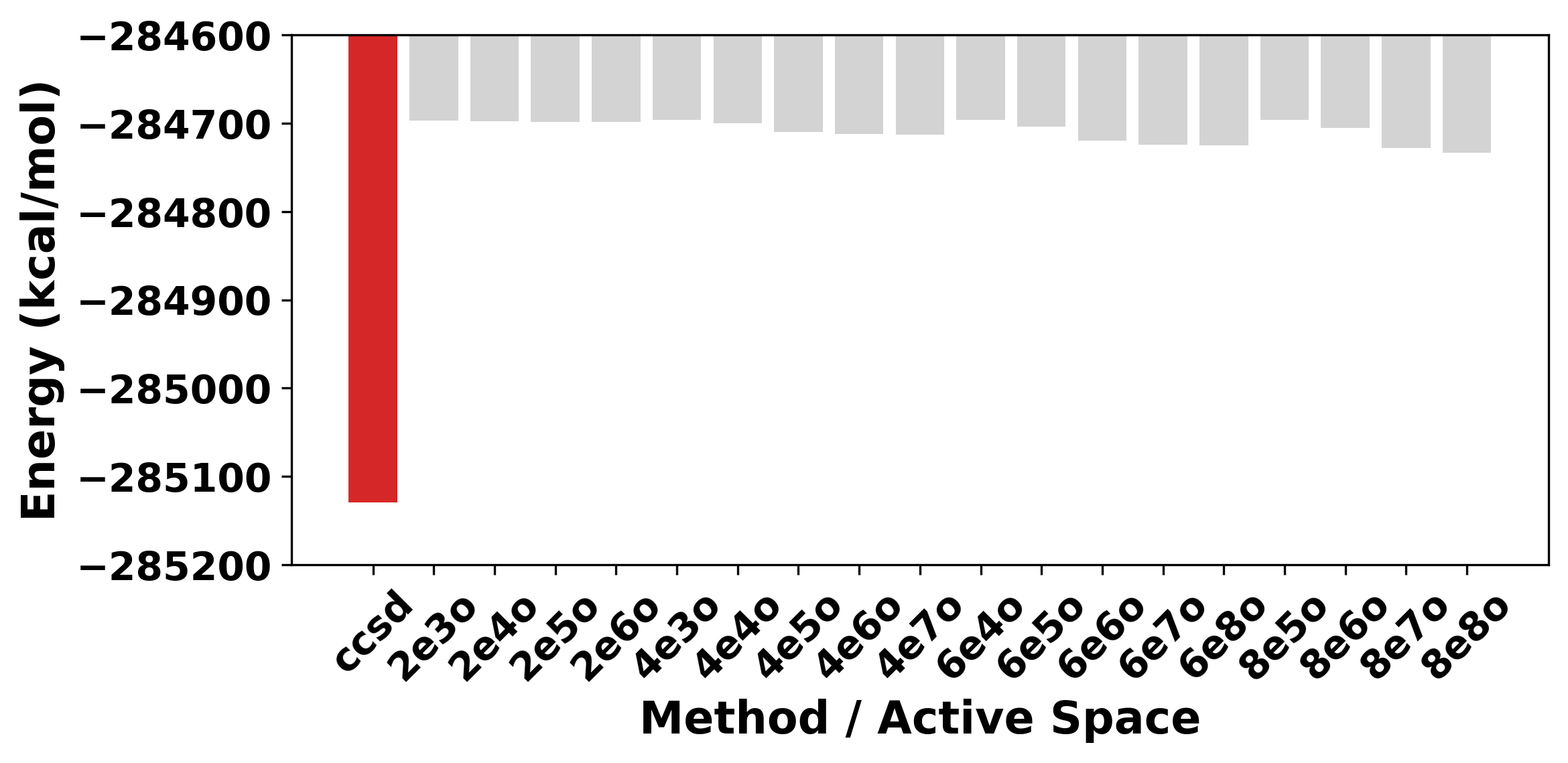}
        \caption{}
        \label{fig:h2_prob}
    \end{subfigure}

    \caption{Comparison of energy values for Alder-ene Reaction a) ene, b) eneophile, c) product, and d) transition state.}
    \label{fig:main}
\end{figure*}

% \FloatBarrier
% \newpage

% \section{Absolute Energy Error and SMF/SMV Comparison for Alder-ene Reaction}

% % %new fig
% \begin{figure*}[h]
%     \centering

%     % ---------- Row 1 ----------
%     \begin{subfigure}{0.4\textwidth}
%         \centering
%         \includegraphics[width=\linewidth]{energy_error_smf_vs_as_ene.png}
%         \caption{}
%         \label{fig:hf_energy}
%     \end{subfigure}
%     % \hfill
%     \begin{subfigure}{0.4\textwidth}
%         \centering
%         \includegraphics[width=\linewidth]{energy_error_smf_vs_as_eneophile.png}
%         \caption{}
%         \label{fig:hf_prob}
%     \end{subfigure}

%     % \vspace{0.3cm}

%     % ---------- Row 2 ----------
%     \begin{subfigure}{0.4\textwidth}
%         \centering
%         \includegraphics[width=\linewidth]{energy_error_smf_vs_as_product2.png}
%         \caption{}
%         \label{fig:h2_energy}
%     \end{subfigure}
% %     %% \hfill
%     \begin{subfigure}{0.4\textwidth}
%         \centering
%         \includegraphics[width=\linewidth]
%         {energy_error_smf_vs_as_ts2.png}
%         \caption{}
%         \label{fig:h2_prob}
%     \end{subfigure}

%     \caption{Absolute Energy Error and SMF/SMV values for (a) ene, (b) eneophile, (c) product, and (d) transition state.}
%     \label{fig:main}
% \end{figure*}

% \begin{figure*}
%     \centering
%     \includegraphics[width=0.5\linewidth]{pes_energy2.png}
%     \caption{Comparison of PES for CCSD and SMF-VQE for Alder-ene Reaction}
%     \label{fig:placeholder}
% \end{figure*}